\documentstyle[prb,aps,psfig]{revtex}
\begin{document}
\draft

\twocolumn[\hsize\textwidth\columnwidth\hsize\csname
@twocolumnfalse\endcsname

\widetext
\title{  Generalized Lanczos Algorithm for Variational Quantum Monte Carlo } 
\author { Sandro Sorella}
\address{ Istituto Nazionale per la  Fisica della Materia and
  International School for Advanced Studies
Via Beirut 4, 34013 Trieste, Italy }
\date{\today}
\maketitle
\begin{abstract}
We show that the standard Lanczos algorithm can be efficiently implemented 
statistically  and self consistently improved, using the stochastic reconfiguration method, 
which has been recently introduced to 
stabilize the  Monte Carlo sign problem instability. With this 
scheme a few  Lanczos steps over a given variational wavefunction are 
possible even for large size 
as a particular case of a more general and more accurate technique 
that  allows to obtain lower variational energies. This method   has been tested 
extensively for a strongly correlated model like the t-J model.
With the standard  Lanczos technique  it is possible  
to compute any kind of correlation functions, with no particular
 computational effort.
By using that the variance $<H^2>-<H>^2$ is zero for an exact 
eigenstate, we show that the approach to the exact solution with few Lanczos iterations is indeed 
possible even for $\sim 100$ electrons  for reasonably good   initial  
wavefunctions. 

The variational stochastic reconfiguration technique  presented here
allows in general   a many-parameter 
energy optimization of any computable many-body wavefunction, including for instance  generic 
long range Jastrow  factors  and  
 arbitrary site dependent orbital determinants. This scheme 
improves  further the accuracy of the  calculation, 
especially  for long distance  correlation functions.
 
\end{abstract}
]

\narrowtext

\section{introduction}

The GFMCSR was introduced recently\cite{sorella} as a method
 to stabilize 
the sign problem instability in the Monte Carlo simulation, for -so called- 
projection techniques, aiming to determine the ground state (GS)  
wavefunction $\psi_0$  of a given Hamiltonian $H$.

In the statistical  approach the electron GS  wavefunction  is   
sampled over
 a set of configurations $\left\{ |x> \right\}$, denoting spins and electron positions 
, and belonging to a complete 
 basis with large or even infinite dimension. 
 The ground state component $\psi_0$  
 of  a given trial state $\psi_G$- henceforth assumed to be non zero on each configuration $x$- 
is filtered out by applying to it some 
projection operator $G^n$ for large number $n$ of power iterations.
The matrix  $G$ maybe given 
by  $G=e^{-H  \Delta t} $ for short imaginary time 
$\Delta t$, as is the case for the conventional diffusion Monte Carlo  
for continuous models, 
or the one that will be considered in the following sections:
\begin{equation} \label{defgreen}
 G=(\Lambda I  -H)^{k_p}
\end{equation}
  for lattice models, where  
  a suitably  large constant shift $\Lambda$ allows convergence 
to the ground state ($\Lambda=0$ is used  for the t-J model considered later on).
The integer  $k_p$, determining the number of powers  of the Hamiltonian 
in the Green function,  
maybe in principle larger than one\cite{caprio}, but 
in the following we avoid this complication and we consider  
$k_p$ fixed at its minimum value: $k_p=1$.  

In order to perform stable simulations with large signal to noise ratio 
even for large  $n$ 
the many-body  propagation
$\psi_n \to G^n |\psi_G>$  
 needs to be   stabilized  at each iteration,  
using an approximation  that can be efficiently implemented statistically 
by means of the stochastic reconfiguration (SR) scheme.\cite{caprio}

The essential step in the SR  is to replace the many-body state  
$\psi_{n+1}(x)=G \psi_n$,  obtained by applying to $\psi_n$  the exact Green function $G$, ` 
 with the approximate state $\psi^\prime_{n+1}(x)$, 
determined by  the following conditions. 
 A given set of $p+1$ operators $\left\{ O^k \right\}$, not restricted to be Hermitian operator, 
   satisfy the following equalities:
\begin{equation}\label{srcon}
<\psi_G|O^k|\psi^\prime_{n+1}>=<\psi_G| O^k|\psi_{n+1}>~~ {\rm  for  }~ k=0,\cdots p  
\end{equation}
where $O^0$ for $k=0$ being the identity here for simpler and more compact  notations,
$\psi^\prime_{n+1}=r_x \psi^f(x)$, $\psi^f$ being a reference state known 
exactly (e.g. the standard variational approach) 
 or statistically (e.g. the fixed node approximate state) 
on each configuration $x$ and 
\begin{eqnarray} 
r_x&=&\sum\limits_{k=0}^p \alpha_k  O^k_x  \label{defrx} \\
{\rm where ~~} O^k_x &=& { <\psi_G|O^k|x>  \over <\psi_G|x> }  \label{okx} 
\end{eqnarray}
and the constants $\alpha_k$  are determined by the conditions (\ref{srcon})
\begin{equation}\label{defalpha} 
\alpha_i =\sum_k s_{i,k}^{-1} <\psi_G| O^k|\psi_{n+1}>
\end{equation}
where the covariance matrix is given by 
\begin{equation} \label{defs}
s_{i,k}=\sum_x \psi_G(x) \psi^f(x) O^i_x O^k_x .
\end{equation}
 
  In this way the SR can be considered a projection $P_{SR}$ of the exactly propagated 
wavefunction    $\psi_{n+1}= G \psi_{n}$  
onto a subspace  spanned by the states $|\psi_f^k>$, defined   
by $<x|\psi_f^k>=O^k_x \psi^f(x)$,
for $k=0,\cdots p$. Notice also that {\em only}  when $\psi^f=\psi_G$, $\psi^k_f= (O^k )^{\dag}| \psi_G>$
,namely $\psi^k_f$ is obtained by applying the operator $(O^k)^{\dag}$ to the state $\psi_G$.

Solving the linear system determined by the SR conditions (\ref{srcon}) we obtain a close 
expression for the linear operator $P_{SR}$, which explicitly depends on the state 
$\psi_G$ and the reference state $\psi^f$:
\begin{equation} \label{defp}
P_{SR}=\sum_{i,j} s^{-1}_{i,j} |\psi^i_f> <\psi_G| O^j 
\end{equation}

The operator $P_{SR}$ is not a true projection operator, 
Though it satisfies the simple requirement (i) $P_{SR}^2=P_{SR}$, 
(ii)  $P_{SR}^\dag = P_{SR}$  is not generally satisfied.
The way to implement statistically this  ''pseudo''-projection operator  in the large 
number of walker limit, was discussed in ref.\cite{caprio}.

After each reconfiguration the state $\psi_n$ is replaced by  $\psi^\prime_n=P_{SR} \, \psi_n$, but also  
the reference state $\psi^f$ is changed. In fact  in the original formulation\cite{sorella,caprio} 
also the reference state explicitly depends on $n$ and is updated 
after each SR:
\begin{equation}  \label{defpsif}
\psi^f_n(x) \to  {\rm sgn}~( \psi_G(x) ) | \psi^\prime_n (x)|.
\end{equation}
The reason of the choice (\ref{defpsif}) 
 is  to optimize the reference state and obtain 
 more accurate results, as explained in the forthcoming paragraphs. 

With the restriction of a stable simulation the signs of the reference
 wavefunction have to be fixed, otherwise the average sign will drop 
exponentially to zero in the simulation, 
Therefore in Eq.(\ref{defpsif})   $\psi^f$ has been  restricted to have 
the same signs (or the same nodes) of the guiding wavefunction $\psi_G$.
On the other hand  the amplitude of the reference wavefunction $\psi^f$
 can be considerably improved  from  our best variational guess $\psi_G$, 
 since during the exact dynamic the state $\psi_n$ gets closer 
to the ground state wavefunction.
The optimal choice for the amplitudes has naturally led to the definition 
(\ref{defpsif}).

This technique, with the choice (\ref{defpsif}), 
 has been shown to be remarkably accurate for  frustrated spin or
 boson systems, allowing  in many test-cases  
  to obtain  essentially  exact results within statistical
 errors.\cite{trumper,j1j2}
However for fermion problems the situation is much different. 
Though this technique  
allows a significant improvement in the energy and correlation 
function calculations\cite{tj,tjnew}; the 
bias remains still sizable and difficult to eliminate completely  
by increasing the number of correlation functions used in the SR technique.
Due to the antisymmetry of the fermion many body wavefunction 
it appears that the nodes in this case play a much more 
important role. 

It is instructive to consider the case of continuous models. 
In this case, for  fermion systems,   nodes  have to appear in 
the many-body wavefunction just 
by anti-symmetry considerations. On the other hand symmetry alone 
does not restrict the nodal surface (the locus  where   the wavefunction 
$\psi_0(x)$ vanishes), implying that   correlation effects can significantly  change the nodal 
surface.
In this case it maybe useless or irrelevant  to improve the amplitudes without 
changing the nodes in the reference wavefunction (\ref{defpsif}). 

For fermion systems, due to the above difficulty to determine 
 a nodal surface which 
is weakly modified  by correlation effects, 
it appears at the moment difficult to avoid a sizable ''nodal'' error 
 in energies and correlation functions. 
Therefore in this case it is extremely important that 
any   approximation is at least controlled by the variational principle.
In this way an approach  different  from the one 
proposed before\cite{sorella}  can be used. The  first step 
to obtain a rigorous variational method is 
to consider  the reference wavefunction $\psi^f$ fixed to the variational or 
the fixed node state.\cite{ceperley1,bemmel}

\section{Fixed reference dynamic} \label{boh}  
In the GFMC propagation a large number $M$ of walkers is used, the $j^{th}$  
  walker is characterized   by two weights $w^f_j,w_j$, 
 acting on a configuration state $x_j$, e.g. a state with definite electron 
positions and spins.  
The reference  weights $w^f_j$ sample statistically the reference state 
$\psi^f_n(x)$ whereas the weights  $w_j$ refers to the 
state $\psi_n(x)$ propagated by the exact Green function 
$\psi_n=G\psi_{n-1}$.
More precisely, taking into account the importance sampling 
transformation\cite{caprio}: 
\begin{eqnarray}
 << w^f_j  \delta_{x,x_j} >> & = & \psi^f_n (x)  \psi_G(x)  \label{defpsifn} \\
 << w_j \delta_{x,x_j} >> & = & \psi_n(x)  \psi_G(x)  \label{defpsi} \\
\end{eqnarray}
where the brackets $<< >>$ indicate 
 both the statistical average and the one over the number of walkers, 
at a given Markov iteration $n$.
The two wavefunctions $\psi^f_n$ and $\psi_n$ are propagated, 
using  the statistical approach.
For the first state   a reference Green function 
$G^f_{x^\prime,x}$ with all positive matrix elements is used, 
whereas the latter one is propagated by means of  the exact Green function 
(\ref{defgreen}) which is related  to the  reference one 
by a simple relation  $G_{x^\prime,x}=s_{x^\prime,x} G^f_{x^\prime,x}$ 
with finite and known  matrix elements  $s_{x^\prime,x}$:
\begin{eqnarray} 
 \psi_{n+1}(x^\prime) \psi_G (x^\prime) & = &  \sum_x G_{x^\prime,x} \psi_{n}(x) \psi_G (x) \label{proppsi} \\
 \psi^f_{n+1}(x^\prime) \psi_G (x^\prime) & = & 
 \sum_x G^f_{x^\prime,x} \psi^f_{n}(x) \psi_G (x)
  \label{proppsif}
\end{eqnarray}
The reference Green function $G^f_{x^\prime,x} =p_{x^\prime,x} b_x$ 
is written in terms of a stochastic matrix $p_{x^\prime,x}$ 
($p_{x^\prime,x} \ge 0$ and $\sum_{x^\prime} p_{x^\prime,x}=1$) times 
an $x$ dependent  normalization  factor 
$b_x=\sum_{x^\prime} G^f_{x^\prime,x}$, so that a statistical 
implementation of the iteration  (\ref{proppsi}) is possible. Namely given the configuration 
$x_j$ of the $j^{th}$ walker a  new configuration $x^\prime_j$ is selected statistically  
with probability $p_{x^\prime_j,x_j}$ (notice $\sum\limits_{x^\prime} p_{x^\prime_j,x_j}=1$ by definition):
\begin{equation} \label{defxp}
 x^\prime_j= x^\prime {\rm ~with ~ probability~~} p_{x^\prime,x_j}.
\end{equation}
 Then the reference weight:
\begin{equation} \label{scalewf}
w^f_j \to b_{x_j} w^f_j
\end{equation}
is scaled by the normalization factor 
$b_{x_j}$, whereas the weight related to the exact Green function $G$,
\begin{equation}  \label{scalew}
w_j \to s_{x^\prime_j,x_j} b_{x_j} w_j  
\end{equation}
is  further multiplied by the $s_{x^\prime_j,x_j}$ matrix element.
  
According to the above Markov iteration 
, defined by Eqs.(\ref{defxp}-\ref{scalew}) all the walkers propagate independently each others, 
The reference weights $w^f$  remain positive, whereas the 
 ones  $w_j$, related to the exact propagation,  accumulate many 
sign changes , leading, for large $n$, to an exponential increase 
of the signal-to noise ratio. 
This is in fact determined 
by the corresponding exponential drop of the average walker sign 
$<s>={  \sum_j w_j \over \sum |w_j| } $  (the infamous  ''sign problem'').
The stochastic reconfiguration allows to alleviate this disease, by implementing 
statistically the operator $P_{SR}$ described  in the previous section.
In practice the weights $w_j$ are replaced by approximate weights $p_j$, but with much  
larger average sign $<s>$.  The weights $p_j$  sample statistically 
$\psi^\prime_n (x) \psi_G (x)$  and are determined 
 by solving the corresponding linear system (\ref{srcon}).
This is done at a given  Markov iteration by averaging Eq.(\ref{srcon}) {\em only} 
 over the walker samples. This  means that 
there exists a bias  due to the  statistical uncertainty of  the quantum averages in 
Eq.(\ref{srcon}).
 This bias however can be controlled efficiently\cite{caprio} 
because for large number of 
walkers it  vanishes as $1/M$.  In this limit 
$r_x$ in Eq.(\ref{defrx})  depends only on the configuration $x$ 
as the statistical fluctuations of the constants $\alpha_k$ can be neglected
 for $M>>p$.

 Another method to control much more efficiently 
the statistical fluctuations of the constants $\left\{ \alpha_k \right\},$ 
  without using a too large number 
of walkers,   is to 
perform the SR scheme  
 only after applying, at each iteration $n$  a  large number $L_b$ of statistical 
steps defined in Eqs.(\ref{defxp}-\ref{scalew})  
 In this   way  
the statistical averages in Eq.(\ref{srcon}) have  only small fluctuations 
$\propto {1\over \sqrt{L_b} }$ that can be  neglected in the limit of large
bin length $L_b$, even when a small number of walkers is used.
In each bin  we are considering only the iteration $n \to n+1$  of the power method when  
the parameters $\alpha_k$ of the wavefunction $P_{SR} \psi_n$ are known and computed in the 
previous iteration. Thus $\left\{ w^f_j \right\}$ evolve 
 statistically  according to $G^f$ (Eqs.\ref{defxp},\ref{scalewf}), and  
new configurations  $\left\{ x_j \right \}$ are generated  for the statistical sampling of 
 $\psi_G(x)~\psi^f(x)= << w^f_j \delta_{x,x_j} >>$, whereas, 
inside the bin,  the weights $\left\{ w_j  \right\}$ are always initialized to   $w_j=r_{x_j} w^f_j$- i.e. 
with the same configurations and a simple scaling of the weights 
it is possible to sample the propagated state  $P_{SR} \psi_n $
 in all the  $L_b$ statistical steps.
Then at each step inside  the bin  
 the exact Green function $G$ 
is applied statistically (Eq.\ref{scalew}) in order to sample $\psi_{n+1}=G P_{SR} \psi_n$, which 
is required to calculate the RHS of the SR conditions (\ref{srcon}). 
At the end of the bin new $\alpha_k$ can be computed by solving the linear system 
(\ref{srcon}). It is understood that even inside the bin a branching scheme at 
fixed number of walkers\cite{calandra} and with a fixed number of correcting factors 
 can  be used  to improve importance sampling.
 Whenever  the length of the bin is so large that the statistical fluctuations can be neglected, 
the algorithm is deterministic inside the statistical uncertainties, and becomes 
much more efficient compared to the original scheme\cite{sorella,caprio}, where the SR conditions
where applied at each step ($L_b=1$). 
We have found instead that  the most efficient scheme is to change these constants $\alpha_k$ only 
when the SR conditions (\ref{srcon}) are not satisfied within statistical errors (e.g. they are 
off by more than three error bars), by increasing systematically the bin length at each iteration.
This scheme allows also  to eliminate the bias due to the finite 
number of walkers  $M$, which was rather sizable in the original formulation\cite{caprio}.


After the SR, say at the iteration $n_0$, 
in order to continue the power method iteration for  $n>n_0$,     
the new approximate state $\psi^\prime_{n_0} = P_{SR}\,  \psi_{n_0}$ replaces $\psi_{n_0}$  but 
the reference state $\psi^f$  is still arbitrary. 
Instead of changing the reference weights with the choice\cite{sorella,caprio}  $w^f_j= |p_j|$, 
(implying Eq.\ref{defpsif}  in the large number of walker limits) 
 it is instead possible  
 to remain with the same  reference state $\psi^f_{n_0}$, without changing 
it in the statistical sense. This is obtained by the following simple scheme.
After the SR new configurations 
$x^\prime_i=x_{j(i)}$ are selected among the old ones $x_j$ 
 according to the probability  $\Pi_j={ |p_j|\over \sum_k |p_k|}$. This  scheme 
 naturally defines  the random index table $j(i)$\cite{calandra,caprio}, used to improve 
importance sampling- as in the branching scheme for the standard diffusion Monte Carlo\cite{nandini}-.
and   allows to continue the simulation more efficiently  with equally weighted walkers 
$|w^\prime_i| \simeq  {\it Const.}$.
In fact  in order to sample statistically the states $\psi^\prime_n$ and $\psi^f$ 
with corresponding new weights $w^\prime_i$ 
and $w^{f^\prime}_i$: 
\begin{eqnarray} \label{reweight}
\psi_n^\prime(x) \psi_G(x) &=&<< p_i \delta_{x,x_i}>>=<< w^\prime_i \delta_{x,x^\prime_i} >> \nonumber \\
\psi^f_n(x)  \psi_G(x) &=& << w^f_i \delta_{x,x_i} >>  =
 << w^{f^\prime}_i \delta_{x,x^\prime_i} >>
\end{eqnarray} 
it is  enough to use  the so called reweighting method, which makes
the above equations  {\em exact} in the statistical sense:
\begin{eqnarray} 
  w^\prime_i &=& {\bar w }~ {\rm Sgn }\, p_{j(i)} \label{reweightw}  \\  
  w^{f^\prime}_i &=& {\bar w}~ w^f_{j(i)}/|p_{j(i)}|={|w^\prime_i|\over |r_{x^\prime_i}| } \label{reweightwf} 
\end{eqnarray}
where $\bar w={1\over M} \sum_j |p_j|$  is a constant common to all 
the walker weights and ${\rm Sgn\, a } =\pm 1$ represents  the sign of the number $a$.
The correcting factor $\bar w$ is taken into account only for a 
finite number $L$ of past iterations, starting e.g. from $n-L+1$, otherwise 
the weights of the walkers may increase or decrease exponentially leading
 to a divergent variance.
This introduces a systematic bias that vanishes however exponentially in 
$L$ and decreases as $1/M$.\cite{caprio}
 In practice for large number of walkers 
it is enough to consider only few (or even none)    ''correcting'' factors 
in the statistical averages, as  common practice 
in Green Function Monte Carlo.\cite{nandini}.
From the reweighting method it is also clear that the choice of the probability 
function $\Pi_j$ is not restricted to be proportional to $p_j$. In particular  
we have found more convenient to use the weights corresponding to $G \psi_n $ 
determined after applying  Eq.(\ref{scalew}), so that the choice  $\Pi_j= {|w_j| \over \sum |w_k| }$ 
further improves importance sampling,  
with  a minor change in Eqs(\ref{reweightw},\ref{reweightwf}).    
\begin{eqnarray*} 
  w^\prime_i &=& {\bar w }~ {p_{j(i)} \over | w_{j(i)} |}    \\  
  w^{f^\prime}_i &=& {|w^\prime_i|\over |r_{x^\prime_i}| }  
\end{eqnarray*}
with $\bar w= {\sum |w_j| \over M}$.  

 Using the previous scheme  the reference state equilibrates necessarily 
to the largest right eigenstate 
of the reference matrix $G^f$.
At equilibrium the state $\psi_{SR}$ 
 has therefore 
a very well compact and clear definition. It represents the maximum right 
eigenstate of the matrix
\begin{equation} \label{srham}  P_{SR}\,  ( \Lambda  I -H )\,   P_{SR},
\end{equation}
with given $\psi^f$   in the definition of $P_{SR}$ (\ref{defp}).

{\em The method is therefore  rigorously variational  provided the pseudo 
projector $P_{SR}$.is
 a true projector operator, namely for $\psi^f=\psi_G$ when $P_{SR}^\dag=P_{SR}$}, as implied 
by Eq.(\ref{defp}). 
In this case in fact by standard linear algebra, 
the maximum right eigenstate of the operator (\ref{srham}) is the best variational state 
of $H$ belonging to the subspace defined by the projector $P_{SR}$i (see Appendix \ref{linear}).

\subsection{Variational energy when  $P_{SR}=P^{\dag}_{SR}$}
 
In the original formulation of the SR  the reference Green function was
defined with a slight generalization of the fixed node Green
 function\cite{sorella}. The Green function  proposed by Hellberg and 
Manusakis\cite{manusakis}  is instead  more appropriate in this context
 and much more convenient from the practical point of view of reducing 
statistical fluctuations: 
\begin{equation} \label{manusakis}
G^f_{x^\prime,x} = { 1 \over z_{x^\prime}} 
| \Lambda \delta_{x^\prime,x} -  \psi_G(x^\prime) H_{x^\prime,x} 
 / \psi_G(x)  |  
\end{equation}
where $z_x=\sum\limits_{x^\prime} 
| \Lambda \delta_{x^\prime,x} -  \psi_G(x^\prime) H_{x^\prime,x} 
/   \psi_G(x)|$, $ | a | $ meaning the absolute value of the number $a$.
It is simple to show that,  by applying the power method with the above 
Green function,  convergence is reached when the maximum right eigenvector 
$ \psi_G(x)^2$ is filtered out:
\begin{equation}
\sum_x G^f_{x^\prime,x} \psi_G(x)^2 = \psi_G(x^\prime)^2
\end{equation}
Thus the above Green function can be used to generate statistically configurations 
( see Eqs.\ref{defxp},\ref{scalewf}) distributed according to $\psi_G(x)^2$  with a stochastic 
matrix $p_{x^\prime,x}= G^f_{x^\prime,x} z_{x^\prime}/z_x$.

The advantage of using the reference Green function (\ref{manusakis})  is evident 
when we consider its {\em very simple}   relation  with  the exact Green function, namely:
\begin{equation} \label{defsxxp}
s_{x^\prime,x} =  G_{x^\prime,x}/G^f_{x^\prime,x}= \pm z_{x^\prime} 
\end{equation}
where the sign  $\pm$ is given by the  sign of the Green function matrix element 
$\Lambda \delta_{x^\prime,x} -  \psi_G(x^\prime) H_{x^\prime,x} /  \psi_G(x) $, 
and depends of course on $x$ and  $x^\prime$. 
Using the fixed reference algorithm in Eq.(\ref{reweightwf}) 
$\psi^f=\psi_G$ and the operator $P_{SR}$ represents 
in this special case a true projector one $P^\dag_{SR}=P_{SR}$.
 Thus in this case the method is rigorously variational as pointed out in the
 last part of the previous section.

We notice also an important property of this method. 
If  in the SR conditions (\ref{srcon})  only operators defined 
by powers of the hamiltonian $O^k=H^k$ are used, the projector $P_{SR}$  acts on 
 the same Krilov basis (spanned by $H^k |\psi_G>$ ,$k=0,\dots,p$) 
of the well known Lanczos algorithm.  
Thus  $(P_{SR}\,  G\,  P_{SR})^n | \psi_G> $ filters out the lowest energy
  variational state in this  Krilov basis, i.e. 
 by definition the state obtained by 
 applying $p$ Lanczos iterations to $\psi_G$.
We recover in particular a known property of the Lanczos algorithm, 
valid also for the SR method: 
the method is exact if $p$ equals the dimension of the Hilbert space.

 Due to the  equivalence of the Lanczos algorithm with the SR technique 
it is  clear why, with  the latter technique, it is possible to obtain rather good
 approximate ground state with $p$ reasonably 
small.\cite{rice,trumper,j1j2,tj,tjnew}  
In fact the convergence of the Lanczos algorithm is 
at least  exponential in $p$.\cite{book}

We have therefore derived that the Lanczos algorithm can be implemented 
statistically using the SR method. This allows  to perform  easily 
two Lanczos iterations on a given variational wavefunction for fairly large
size systems.
Furthermore the SR method allows to put several correlation functions in the (\ref{srcon}). Since the 
method is strictly variational,   
 the variational energy has necessarily  to decrease by increasing the mentioned 
 number of correlation functions.

\subsection{ Improving the variational energy } 
Following \cite{tklee}, by applying a finite
number of exact Green function iterations  to  the wavefunction $\psi_{SR}$,
the corresponding quantum average, 
\begin{equation} \label{defpowers} 
E_{SR}^k= { <\psi_{SR}|G^k H G^k| \psi_{SR}> \over < \psi_{SR} | G^{2k}
 |\psi_{SR}> } 
\end{equation} 
remains obviously variational for any $k$.

 Taking into account $2 k$  statistical factors $s_{x^\prime,x}$, the 
above quantum averages can be statistically evaluated with the same Markov chain 
for which $E_{SR}$ ($k=0$) is computed.

 The sign problem can be controlled for not too large $k$ and  systematically 
improved variational energies  can be obtained compared to the $k=0$ result.
However experience has shown that it is very difficult to have significant improvement
over the $k=0$ result for large system size.

\section{Variational energy when $P_{SR}\ne P^{\dag}_{SR}$}

We have seen that by changing the definition of the reference weight 
after the SR was applied 
(\ref{reweightwf}) the method is rigorously variational. However , as we 
have explained in the introduction,   a  better  choice  is to continue
 after the SR with
$w^{f^\prime}=|w^\prime|$.
The rational of this choice being that
the wavefunction $\psi^\prime=P_{SR}\,  \psi $ has 
much better amplitudes than the variational wavefunction $\psi_G$. 
This allows to improve self-consistently the reference wavefunction
 in order  to be as close as possible to the  true ground state.
In this way however $\psi^f \ne \psi_G$ and the method is no longer 
variational in the sense that the SR state defined by the right eigenstate
with maximum modulus $| \Lambda-E_{SR}|$ eigenvalue of the matrix :
\begin{equation}
 P_{SR} \, (\Lambda I -H )\, P_{SR} | \psi_{SR}> =(\Lambda - E_{SR})| \psi_{SR}>
\end{equation}
is no longer the lowest energy one in the basis defined by $|\psi_f^k>$  
,$k=0,1,\dots,p$ and $E_{SR}$ does not necessarily bound the ground state 
energy.

 A compromise between the two methods, is to introduce a parameter $r$
 that interpolates smoothly the 
two limits. This parameter enters in the reweighting relation (\ref{reweightwf}) in a very simple manner:
\begin{equation} 
  w^{f^\prime}_i = ={|w^\prime_i|\over |r_{x^\prime_i}|^{1-r} }  \label{defgamma}
\end{equation}
Notice that, when we release the fixed reference dynamic for $r \ne 0$ and when for large $n$ the 
constants $\left\{ \alpha_k \right\}$ converge to non zero values, 
 the reference wavefunction $\psi^f$ is not given by averaging the configurations 
with the weights $w^f$ since the relation $\psi_{SR}=P_{SR} \psi_n(x) =r_x \psi^f (x)$  is implied 
by  the definition of the projector $P_{SR}$ in Eq.(\ref{defp}). 
In fact,  by  using Eq.(\ref{defgamma}) and that
 $<< w^\prime_i \delta_{x,x_i} >> =\psi_{SR}(x) \psi_G(x) =r_x \psi^f(x) \psi_G (x)$, it is simple to derive
 that:
\begin{eqnarray} \label{relpsif}
<< w^{f^\prime}_i \delta_{x,x_i} >> &=& |r_x|^{r-1} {\rm Sgn}\,  r_x  << w^\prime_i \delta_{x,x_i} >> \nonumber \\
&=& |r_x|^{r-1} {\rm Sgn}\, r_x \, \psi_{SR}(x) \psi_G (x) \nonumber \\
&=& |r_x|^{r}  \psi^f (x) \psi_G (x)
\end{eqnarray}

 For $r=0$ the method is rigorously variational, in the sense described before. It is 
reasonable to expect a  similar behavior  for $r << 1$ when $P^{\dag}_{SR}\simeq P_{SR}$, 
$r=1$ being the optimal choice for the amplitudes of the reference wavefunction $\psi^f$.
 
For $r \ne 0$ we  assume that 
the infinite number of walkers or large bin $L_b$ is taken so that the parameters
 $\alpha_k$ in the definition of 
$r_x$ (\ref{defrx}) can be considered constants  for large $n$.
In this case, if we take into account the reweighting (\ref{defgamma}), the reference
 wavefunction $\psi^f$ is obtained as the  right eigenvector  
$\psi_R(x)=<< w^{f^\prime}_i \delta_{x,x_i} >>$  with maximum eigenvalue of 
the renormalized reference Green function $\bar G^f$:
\begin{equation} 
\bar G^f_{x^\prime,x} = | r_{x^\prime}  |^r G^f_{x^\prime,x}   \label{defgrefb}
\end{equation}
,namely as $\psi_f(x)=|r_x|^{-r} \psi_R(x)/\psi_G(x)$ 

 Then the $SR$ state $\psi_{SR}(x)=r_x \psi^f(x)$ 
is simply given in terms of this right eigenvector:
\begin{eqnarray}
\psi_{SR}(x)&=& R_x \psi_R(x) \label{defpsisr}  \\
{\rm with } ~~ R_x&=& |r_x|^{1-r}/\psi_G(x) 
~ {\rm Sgn}\, r_x  \label{defRx} 
\end{eqnarray}
Thus even when $r \ne 0$ the  $SR$  state can be uniquely determined.
It is also clear that, since $r_x$ is not necessarily positive or negative, the nodes can 
be changed and improved with respect to the nodes of the initial guess $\psi_G$, both for $r=0$ with  
the standard Lanczos algorithm and for $r>0$.
If the nodes of  $\psi_G$ are exact and the hamiltonian is not frustrated it is also possible 
to show that $r=1$ and $\Lambda \to \infty$ provide the exact result, with $r_x$ positive definite.

{\em It is possible to compute correlation functions over $\psi_{SR}$? }

\begin{figure}
\centerline{\psfig{figure=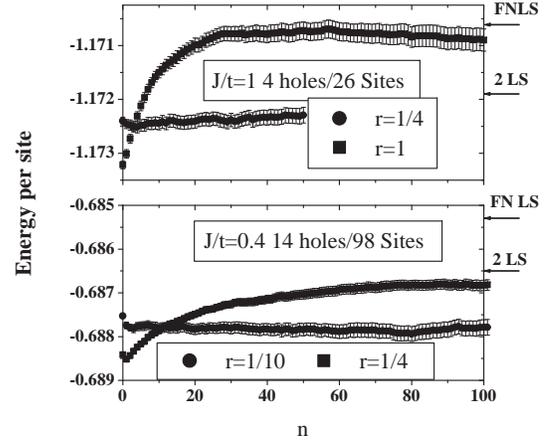,width=8cm}}
\caption{  Energy per site as a function of the forward iteration $n$ as described in the 
 App.(\ref{forwardapp}) for the 26 sites 4 holes case and the 98 site 14 hole case.
The value of the variational parameter $r$ is also indicated.}
\label{forwardfig}
\end{figure}

As it is shown in App.(\ref{forwardapp}) the answer is yes,  and  not only for  correlation functions diagonal in the 
basis $x$ , as in the standard forward walking technique\cite{calandra}, 
but also for all the ones with off diagonal elements contained in the non-zero hamiltonian 
matrix elements.
In particular it is possible to compute :
$$ <\psi_{SR}| H | \psi_{SR}>  \ge E_0$$
i.e. the expectation value over the state defined by the SR conditions.
This estimate is obviously variational and can be further improved, 
 by applying a finite 
number of exact Green function powers to the right and to the left of the hamiltonian
,as in the power-Lanczos algorithm\cite{tklee}, with the difference that in this case 
Eq.(\ref{defpowers}) has to be evaluated with the ''forward walking technique'', as described in the 
App.(\ref{forwardapp}).
In Fig.(\ref{forwardfig}) we plot the evolution of the expectation value of the energy 
over the state $\psi_{SR}$ as a function of the number of  iterations $n$, required by the 
forward walking to filter out from $\psi_G$ its component over $\psi_{SR}$, leading to a true 
variational energy estimate. We see that for $r=1$, within the original SR technique\cite{sorella,caprio}, 
the energy expectation value can be much higher than the corresponding ''mixed average'' estimate 
($n=0$). 

 This behavior  can be understood in the following way   
, for $r=1$ and  for $\Lambda \to \infty$  there is no way to improve the sign 
of the wavefunction over $\psi_G$ because $r_x \to 1$ ($G$ and $G^f$ tend to the identity up to a constant 
and so the correction $r_x$ to $\psi^f \simeq  \psi$  for $r=1$ 
has to became unity up to $O({1 \over \Lambda})$), whereas for $r=0$ 
the Lanczos algorithm , which is $\Lambda$ independent, certainly modifies  and improves 
the nodes. For fermion systems therefore it appears important to work with small $r$ because the nodes of 
the wavefunction play a particular important role in determining a good variational energy.

\begin{figure}
\centerline{\psfig{figure=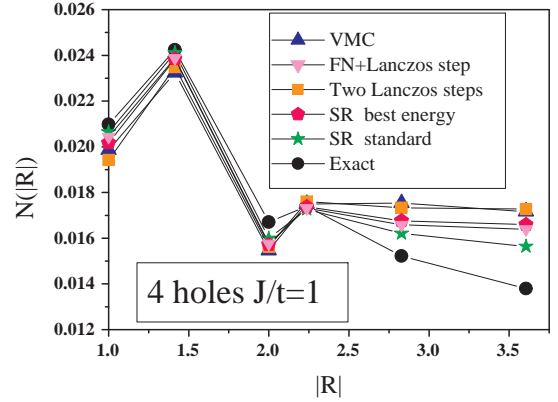,width=8cm}}
\caption{ Hole-hole correlations for the 26 site $t-J$ cluster for various methods. 
''SR standard'' indicates the original GFMCSR implementation ($r=1$ here)\cite{sorella,caprio}, 
 ,whereas ''SR best energy'' indicates the optimal variational SR-wavefunction obtained with $r=0.25$.
 The ''VMC '' is the lowest energy  Jastrow-Slater variational singlet wavefunction as discussed in 
Sec.(\ref{full}), FN+Lanczos step the fixed node over the one Lanczos step wavefunction. 
 }\label{nqj1}
\end{figure}

A different behavior is seen for correlation functions.
 For large $J/t=1$ when a four particle bound state is 
likely to be formed, our Jastrow-Slater wavefunction is not appropriate enough, and the large 
distance behavior is not exactly reproduced (see   Fig.\ref{nqj1}) even when 
we apply to it a couple of Lanczos iterations, that, remarkably,  provide a very accurate variational energy (see table). 

That the qualitative behavior of this correlation function is different from the variational 
starting wavefunction, can be understood only when the algorithm with $r>0$ or the FN (which is worse in energy) 
are  used. Especially 
successfull is the original  technique $r=1$, which improves by a factor of two the important 
long range behavior of $N(|R|)$, clearly displaying the features  of a genuine bound state, by 
a decaying  probability to have holes at larger and larger distances.
For correlation functions diagonal in the chosen basis the nodes do not play any role 
and $r=1$ or the FN itself, are  likely to provide much better correlation functions. However from the 
previous argument about the impossibility to correct the nodes for $r=1$ and  $G, G^f \sim I $ we expect indeed  
that for large size the Green function $G$ tends to the identity, either 
because $\Lambda \propto L$, as required by  the   power method to converge, or because the gap 
to the first excited state decreases  and a power iteration $\psi_{n+1}=G \psi_n$ is  
less and less effective for changing the wavefunction. 
 Thus $r$ has to scale to zero for large size if we do not want to spoil too much 
the variational expectation value of the energy (see Fig.\ref{forwardfig}).
It is remarkable that the gain in variational energy is larger and larger when the size is increased.
Thus the $r>0$ technique seems to overcome, at least partially, 
 a serious limitation of the Lanczos algorithm, namely that in the thermodynamic limit the energy 
per site cannot be improved by a  technique which is not size consistent. 
The gain in energy with $r>0$ can instead be size consistent, since, as shown in the appendix (\ref{forwardapp}),
$r>0$ corresponds to modify the reference Green function $G^f \to \bar G^f$, similarly to what the Fixed node 
algorithm-which is size consistent for size consistent $\psi_G$- does.

\section{ Numerical implementation  }
In order to  put efficiently a finite number $p$ of hamiltonian powers in 
the  SR scheme 
it is by far more convenient to use an importance sampling strategy (see e.g. App.\ref{onelan}), 
 by using information 
of the previous $p-1$ Lanczos iterations. A guiding wavefunction 
corresponding to the previous ($p-1$)  approximation 
$\psi^\prime_G(x)= <x|\sum\limits_{k=0}^{p-1}  \alpha_k H^k |
 \psi_G>=r_{p-1}(x) \psi_G(x)$ can be used.\cite{notezero}
With this new guiding function the powers of the hamiltonian can be put 
in the SR conditions by 
computing corresponding mixed average estimators:
\begin{equation} \label{maused}
O^k_x={ <\psi_G|H^k |x> \over <\psi^\prime_G| x> }
\end{equation}
for $k=0,\cdots p$.

The advantage of using the SR scheme is clear even when we restrict this 
method only to the evaluation of the first few Lanczos iterations. 
In order to perform $p$ Lanczos iterations, it is 
enough to compute only $p$ hamiltonian powers on a given configuration $x$. 
In the conventional variational method  it is always necessary to compute 
the expectation value of the hamiltonian amounting to $2 p+1$ 
powers of the  hamiltonian, leading  to a much more demanding 
numerical effort.
 It is also important to emphasize that within this technique  
 the parameters $\alpha_k$ defining the SR state,  
 are given  
  at the end of the SR simulation $n \to \infty$.
Once the $\left\{\alpha_k \right\}$ are determined it is then convenient to compute 
correlation functions with fixed constants $\left\{\alpha_k \right\}$,  
by performing statistical averages over a large bin $L_b$ without applying the SR 
conditions (\ref{srcon}) as discussed in Sec.(\ref{boh}).
This method significantly improves the statistical fluctuations of the quantum averages over 
the  variational state $P_{SR}| \psi_G>$.

\section{ Energy optimization  for a  many parameter variational  wavefunction} \label{full}
The most important advantage of the SR technique is that many variational parameters 
can be handled at  a little expense of solving a linear system (\ref{defalpha})
 of corresponding size.
The few Lanczos step technique, as we have discussed before, is determined only by  
few coefficients 
due to the difficulty for large size to compute  
 many powers of the hamiltonian on a given configuration $x$,
which is required for the  evaluation of the corresponding mixed estimator 
$O^k_x={<\psi_G|H^k|x>\over <\psi_G|x> }$.

Clearly the method discussed in Sec.(\ref{boh}) is not limited only to the hamiltonian momenta  
correlation functions, but remains variational for arbitrary ones.
In particular many kind of correlation functions  can be thought to be a renormalization 
of the guiding wavefunction $\psi_G$,  
 allowing a powerful multi-parameter energy optimization similar to \cite{umrigar},
 where the variance was instead minimized. In the present chapter we assume for simplicity 
that {\em all} the correlation functions are used for the purpose of optimizing the 
variational wavefunction $\psi_G$, and we restrict 
our very general analysis  to variational wavefunctions of
the Jastrow-Slater form for a strongly correlated system such as 
the t-J model.\cite{anderson,zhang} 
\subsection{The variational wavefunction}
The Jastrow-Slater  variational wavefunction can be generally written as:   
\begin{equation} \label{bcs}
|\psi_G \rangle= \hat{J} |S\rangle.
\end{equation}
where $|S\rangle$ is a determinant  wavefunction,  
that can be obtained as an exact ground state of  a generic one-body  hamiltonian 
of the bilinear form\cite{onebody} $H_{\it 1-BODY} \simeq  c^\dag c \dots + c^\dag c^\dag \dots + {\it h.c.}$  
whereas the Jastrow factor
\begin{equation} \label{Jastrow} 
\hat{J} = exp(\sum_{i,j} v(i,j) n_i n_j ) exp(\sum_{i,j} v^{z}(i,j) \sigma^z_i \sigma^z_j ) 
\end{equation}
introduces arbitrary Gaussian correlations between local 
charges $n_i=\sum_\sigma c^{\dag}_{i,\sigma}.c_{i,\sigma}$ 
and local spins along the z-axis:
$\sigma^z_i=c^\dag_{i,\uparrow}.c_{i,\uparrow} - c^\dag_{i,\downarrow}.c_{i,\downarrow}$ 
Both operators are defined on each configuration $\left\{ x \right\}$ of the chosen Hilbert space.
The reason of the exponential form  in the Jastrow wavefunction comes from size-consistency 
considerations, implying that for macroscopic and disconnected regions of space A and B  
the wavefunction factorizes $\psi_{A,B}=\psi_{A} \bigotimes \psi_B $,
this factorization being fulfilled by the exponential form.

Considering  the $t-J$ model the restriction of no doubly occupied sites
can be thought of an infinitely negative $v(i,i)=-\infty$ charge correlation, whereas restriction 
at fixed number $N$ of particles in a finite size $L$, can be thought of another singular 
Jastrow term $exp(-\infty (N_{tot}-N)^2)$
where $N_{tot}=\sum_i n_i$ is the total charge operator.
The latter two singular terms in the Jastrow factor are  simply  a restriction on the  Hilbert
space that can be very easily implemented without numerical instabilities.

We now consider how the symmetries of the finite size t-J model drastically restrict the 
still too large number of variational parameters in  the Jastrow-Slater wavefunction.
Translation invariance implies that the Jastrow potential depends only on the distance 
$v(i,j)=v(|R_i-R_j|)$, whereas $v^z=0$ for a singlet wavefunction.

The singlet and translation symmetry  imply also  strong restrictions to the one-body hamiltonian.
defining the Slater determinant.
This hamiltonian  can be generally written: 
\begin{eqnarray}
\hat{H}_{S}&=&\hat{H}_0+  (\hat{\Delta}^\dag +\hat{\Delta})  \label{hbcs} \\
\hat{\Delta}^\dag&=& \sum_{\langle R,\vec \tau \rangle} \Delta (\vec \tau)  \big( 
 \tilde{c}^\dag_{R,\uparrow} \tilde{c}^{\dag}_{R+\vec\tau,\downarrow} + 
 \tilde{c}^\dag_{R+\vec \tau,\uparrow} \tilde{c}^{\dag}_{R,\downarrow}  \big)\label{ddag} 
\end{eqnarray}
where $\hat{H}_0=\sum\limits_{k,\sigma} \epsilon_k \, 
\tilde{c}^\dag_{k,\sigma} \tilde{c}_{k,\sigma}$ 
is the free electron tight binding nearest-neighbor Hamiltonian, 
$\epsilon_k=-2 t (\cos k_x + \cos k_y) -\mu$, $\mu$ is the free-electron 
chemical potential and  
$\hat{\Delta}^\dag$ creates all possible pairs at the various distances $|\tau|$  
with  definite rotation-reflection  symmetry (e.g. $d_{x^2-y^2}$ implies
 $\Delta(1,0)=-\Delta(0,1)$)

For a generic Jastrow-Slater singlet state, satisfying all symmetries of the t-J model,
a quite large number of variational parameters are therefore available 
corresponding to  $v(|\tau|)$ and  $ D(\tau)$ for all distances $|\tau|$. 
Not all these  parameters are independent, namely  the substitution 
$v(|\tau|) \to v(|\tau|) +{\it const.}$ 
does not change the wavefunction up to a constant, so that $v(|\tau|)$ can be assumed to be zero at the maximum 
distance. Analogous dependence exists  between the various parameters $\Delta(\vec \tau)$, 
since  after projecting it at fixed number of particles, the ground state 
of the Slater determinant hamiltonian (\ref{hbcs}) 
 can be written:
\begin{equation}\label{wfbcs}
|S>=(\sum_{R,\tau} f(\vec \tau)   \big( 
 \tilde{c}^\dag_{R,\uparrow} \tilde{c}^{\dag}_{R+\vec \tau,\downarrow} + 
 \tilde{c}^\dag_{R+\vec \tau,\uparrow} \tilde{c}^{\dag}_{R,\downarrow} \big)  )^{N/2}  |0>
\end{equation}
where $f(\tau)$ is the pairing wavefunction simply related to $\Delta(\tau)$ in Fourier transform:
$$f(k)={\Delta(k)\over \epsilon_k +\sqrt{\Delta_k^2+\epsilon_k^2} }.$$
where $\epsilon_k$ in the above expression  is not limited to nearest neighbor hopping.

Thus if we scale $f(\tau) \to {\it const.}  f(\tau)$ the many body wavefunction (\ref{wfbcs}) 
 remains unchanged,
 implying that the number of independent parameters is equal  to $N_{shell}-1$, where $N_{shell}$ is 
the number of independent shells  at $|\tau|>0$  consistent 
with the rotation-reflection symmetry of $f$. 
Notice that, once in the determinant part of the wavefunction $N_{shell}-1$  
 variational parameters are independently varied, it  is useless  to consider other terms, such as, for instance, the chemical 
potential $\mu$ or next-nearest neighbor hopping in $H_0$: they always provide a suitable 
renormalization of $f$ that can be sampled by the first $N_{shell}-1$ parameters.
Moreover  by performing a particle-hole transformation in Eq.(\ref{hbcs}) 
on the 
spin down $\tilde{c}^\dag_{i,\downarrow} \to (-1)^i \tilde{c}_{i,\downarrow}$, 
the ground state of the Hamiltonian (\ref{hbcs})  is just a Slater-determinant with 
$N=L$ particles \cite{shiba}. This is the reason why this variational wavefunction
 represents a generic Jastrow-Slater state, a standard 
variational wavefunction used in QMC. Using
the particle-hole transformation, it is also possible to control exactly
the spurious finite system divergences related to the nodes of the 
d-wave order parameter.

\subsection{Stochastic minimization}
Among the correlation functions important to define the variational  
wavefunction  two classes are important for guiding functions 
of the Jastrow-Slater form (\ref{bcs}).
\begin{itemize}
\item the first class of operators renormalize the Slater determinant 
and have been identified by Filippi and Fahy\cite{filippi}, $O^k$ are defined by means of 
 one body operators  $O^k_{\it 1-BODY}$ 
 by the following relation:
\begin{equation}\label{onebody}
O^k_x={ <x|O^k|\psi_G>  \over <x|\psi_G> } ={ <x| O^k_{\it 1-BODY}|S> \over <x|S> }
\end{equation}
Thus for $|\alpha_k|$ small, 
$(1 + \sum_k \alpha_k  O^k) |\psi_G>
\simeq \hat J \times exp(\sum_k \alpha_k  O^k_{\it 1-BODY}) |S>$
which remains a Jastrow Slater wavefunction of the same form $J  S^\prime$ with 
$S^\prime=exp(\sum_k \alpha_k  O^k_{\it 1-BODY}) |S>$
Since one body operators are bilinear (e.g. $c^\dag c$) in fermion second-quantization  operators`
$S^\prime$ remains a Slater determinant.\cite{onebody}
 In the Jastrow-Slater case for the $t-J$ model considered here the one-body operators read:
\begin{equation}
O^k_{\it 1-BODY} = \\
\sum_{R,\tau \in {\it Shell \# k}} S(\tau) \big( 
 \tilde{c}^\dag_{R,\uparrow} \tilde{c}^{\dag}_{R+\tau,\downarrow} + 
 \tilde{c}^\dag_{R+\tau,\uparrow} \tilde{c}^{\dag}_{R,\downarrow}  \big) 
\end{equation}
where the sign $S(\tau)=\pm 1$ is determined by symmetry. 
Also the bar kinetic energy $H_0$ is considered in this approach.  According 
to the previous discussion the chemical potential $\mu$ 
is  fixed to the free electron one in $H_0$. 
\item
the second class of  correlation functions are the ones that appear in the Jastrow factor. They 
are the diagonal operators $O^k$ -- density-density $\sum_R n_R n_{R+\tau}$  or spin-spin 
$\sum_R \sigma^z_R \sigma^z_{R+\tau}$ -- in the chosen basis $x$ 
of configurations with fixed spins and electron positions. 
Again for 
small $\alpha$ they can be considered a renormalization of the Jastrow factor: 
$ J \to J exp(\sum_k \alpha_k O^k)$.
\end{itemize}

The multi-parameter minimization method can be summarized as follows:
\begin{itemize}
\item after each reconfiguration the factor $r_x=\sum_k \alpha_k O^k_x$ is  computed with 
given $\alpha_k$, whose statistical fluctuations can be arbitrary reduced by increasing 
the number of walkers or the bin length $L_b$ as described in Sec.(\ref{boh}).
 In this case of non linear optimization the bin technique 
is particular important  
 because it allows to avoid, for large  enough bin length $L_b$,  unphysical fluctuations of 
the guiding wavefunction. 

After an exact Green function step a wavefunction  better than $\psi_G$ is obtained and 
is parametrized 
by the coefficient $\alpha_k$  contained in the factor $r_x$:
In fact 
$$P_{SR} \psi_{n+1} = P_{SR} G P_{SR} \psi_n= r_x \psi_G(x)$$
has to be by definition a variational state better than $P_{SR}\psi_n$\cite{ineq}, which in turn is 
better than $\psi_G$, since for instance we can assume $\psi_{n=0}=\psi_G$.

\item It is therefore convenient to change at each iteration $n$ the guiding wavefunction:
 \begin{equation} \label{changepsig}
  |\psi^\prime_G> \to exp(\sum\limits_{k=1}^p \bar \alpha_k  O^k)  |\psi_G>
 \end{equation}
In the above equation we have introduced new scaled  coefficients 
$\bar  \alpha_k= \alpha_k / C$ simply related 
to the ones $\alpha_k$ defined by the SR conditions (\ref{srcon}).
This is obtained by recasting $r_x$ in a form that is  more suitable for exponentiation:
\begin{equation} \label{newrx}
r_x=C \left[ 1 +\sum\limits_{k=1}^p \bar \alpha_k (O^k_x -\bar O^k) \right] 
\end{equation}
where 
$$\bar O^k= {\sum_x \psi_G(x)^2 O^k_x \over \sum_x \psi_G(x)^2 }= << w^f_j  O^k_x \delta{x,x_j} >>$$ 
and $C=1+\sum_{k=1}^p \alpha_k \bar O^k$.
The above exponentiation is justified, provided ${ \psi^\prime_G \over ||\psi^\prime_G ||} \simeq 
  {   \psi_G \over || \psi_G ||}$.  This  is certainly fulfilled at equilibrium  
when  for large 
$n$ $\psi^\prime_{n+1}=P_{SR} \psi_{n+1}\propto P_{SR} \psi_n\propto \psi_G^*$, 
$\psi_G^*$ being the Jastrow-Slater wavefunction with lowest energy expectation value.
In fact at equilibrium 
 the SR conditions turn exactly to the Euler equations of minimum energy 
for $\psi_G^*$:
\begin{equation}\label{euler}
{ <\psi_G^*| O^k |\psi_G^*> \over 
<\psi_G^*| \psi_G^*> }
=      {   <\psi_G^*| O^k H |\psi_G^*>  \over 
<\psi_G^* | H  |\psi_G^*> } 
\end{equation}
as implied by (\ref{srcon})  for $\psi^\prime_{n+1}=\psi^*_G$ and 
$\psi_{n+1} =G\psi^\prime_n \propto G \psi^*_G$, and  taking also into account that 
here for simplicity $O^0$ is the identity.

This implies that $\psi^*_G$ is a lowest energy wavefunction of the Jastrow-Slater form. 
There maybe many local minima when the Euler's equations are identically satisfied. In this case 
the SR represents a very useful tool for global minimization. 
In fact the bin length $L_b$ or the number of walkers $M$  represent 
at each iteration $n$ an effective inverse temperature  that can be increased gradually 
following  the well known ''simulated annealing'' statistical algorithm.\cite{devecchi}
As it is shown in Fig.(\ref{melt}) we apply this technique with very short bin length 
using a full translation invariant singlet wavefunction  and 
using x,y reflection symmetries  without rotation symmetry. This 
amounts to 24 independent parameters for a 26 site cluster ($N_{shell}=13$).
In the plot we show the evolution of the short range BCS  parameters $\Delta(1,0),\Delta(0,1)$ 
,when at the beginning $\psi_{n=0}=\psi_G$ was set with
no Jastrow term  $v=0$ and the s-wave symmetric  determinant defined by 
  $\Delta(1,0)=\Delta(0,1)=0.1t$, 
$\Delta_{\tau}=0$ for $|\tau|>1$.
The s-wave symmetric solution is locally stable, but for short enough bin there is 
a finite tunneling probability to  cross the barrier and stabilize the much lower energy 
solution with d-wave symmetry.
\begin{figure}
\centerline{\psfig{figure=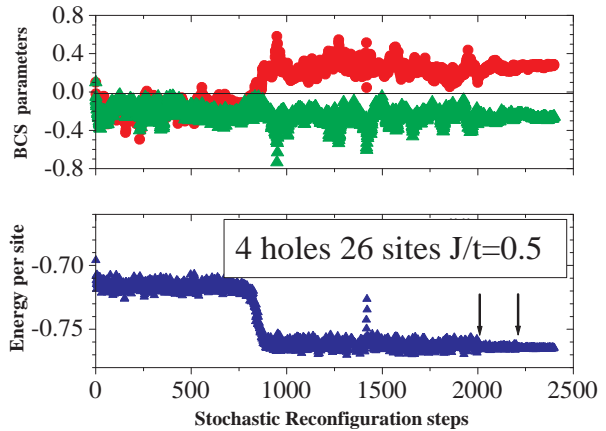,width=9cm}}
\caption{Energy per site and evolution of the pairing amplitudes
for  4 holes in the t-J model at $J/t=0.5$ as a function of the stochastic power 
method iteration $n$. Here 200 walkers were used.
Upper panel: triangles and circles denote 
nearest neighbor  $\Delta(\tau)$  amplitudes in Eq.(\ref{hbcs}) along x and y directions respectively.
Lower panel: the arrows indicate the power iterations when the Monte Carlo bin length $L_b$ 
has increased from 10 to 100 (left arrow) and from 100 to 
500 (right arrow) Monte Carlo steps. }\label{melt}
\end{figure}

\item In order to continue with the new guiding wavefunction (\ref{changepsig}) 
 without another long equilibration,
walker weights $w_j$ and $w^f_j$ in Eqs.(\ref{defpsifn},\ref{defpsi}) can be reweighted
 as follows:
\begin{eqnarray} \label{scalewwf}
w_j&\to & w_j (\psi^\prime_G(x_j)/\psi_G(x_j)) \nonumber \\
w^f_j&\to & w^f_j (\psi^\prime_G(x_j)/\psi_G(x_j))^2 \nonumber \\
\end{eqnarray}
in order that the new weights  acting on the same configurations $x$ 
 represent statistically  Eq,(\ref{defpsifn}) with $\psi^f=\psi_G=\psi_G^\prime$ and Eq.(\ref{defpsi}) 
with the new guiding wavefunction $\psi^\prime_G$.
\item  For large $n$ the one-body operators corresponding to the Slater-determinant 
may became  linearly dependent because $D$ may approach  an eigenstate of a one body 
hamiltonian $\sum h_k  O^k_{\it 1-BODY}$, $\sum h_k   O^k_{\it 1-BODY}  |S>= E |S>$,
 with suitable constants $h_k$ and $E$.  
Thus the covariance matrix $s_{k,k^\prime}$ quickly become singular, leading to a numerical 
instability which is difficult to control statistically.
A stable method to overcome this difficulty was found by  Filippi and Fahy\cite{filippi}, 
 essentially by  
 taking out one operator (the one body kinetic energy) from the ones used in the linear 
system (\ref{srcon}).
 Here we have found more convenient to solve the linear system (\ref{srcon}):
$$ \sum s_{k,k^\prime} \alpha_{k^\prime}  = <\psi_G| O^k G |\psi_n> =f_k$$
by diagonalizing first the symmetric matrix 
\begin{equation} 
s_{k,k^\prime}= \sum_j U_{k,j} \Lambda_j U_{k^\prime,j} 
\end{equation}
 and taking out 
the lowest eigenvalue $\Lambda_0\ge 0$ component 
of the positive definite matrix $s$.
 This  is equivalent to select in the SR 
another set of $p-1$ operators which are no longer singularly dependent.
This is a perfectly legal operation for $\Lambda_0 \sim 0$, since  
as shown in \cite{caprio}, for the singular operator $O^* =\sum_k U_{k,0} O^k$, such 
that $O^* | \psi_G \rangle =0$, the SR condition (\ref{srcon}) is identically satisfied.  
Thus  the resulting linear system is no longer   affected by the above numerical instability:
\begin{equation} \label{related}
 \Lambda_j  \alpha^\prime_j= \sum_k U_{k,j} f_k 
\end{equation}
with $\alpha^\prime_0=0$ and 
\begin{equation}
\alpha_k=\sum_{j>0}  U_{k^\prime,j} \alpha^\prime_j
\end{equation}
 Finally  we obtain a much more stable determination 
of $\alpha_k$, which does not affect the result at the equilibrium, where $\Lambda_0 \to 0$ and 
the correct Euler equations are satisfied.
 With this scheme also the optimization of the Jastrow parameters together with the 
Slater determinant ones is possible without too much effort.  

\item We have found that the generic situation for Jastrow-Slater wavefunctions is  that 
the optimal determinant is actually the ground state of a one body hamiltonian
$H_{1-BODY}=\sum_k h_k  O^k_{\it 1-BODY}$, a  particular 
linear combination of the  chosen one body operators used in the SR conditions. This is 
 in agreement with the Filippi-Fahy  
ansatz\cite{filippi}.  Occasionally however, the optimal determinant turns out to be 
an excited state of a one body hamiltonian.
\end{itemize}

\begin{figure}
\centerline{\psfig{figure=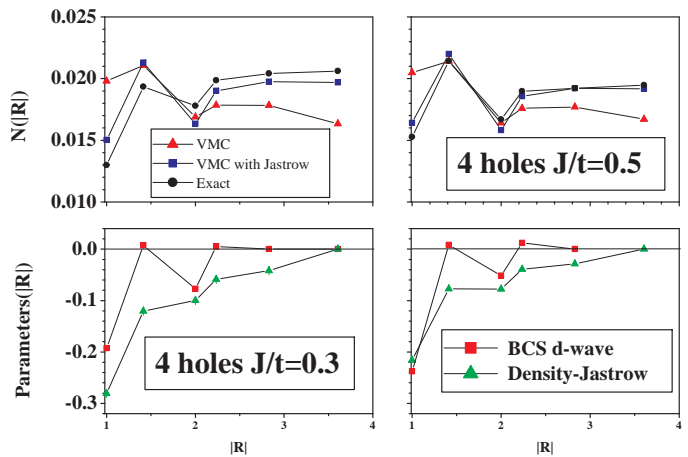,width=10cm}}
\caption{Hole-hole correlations (upper panels) and Jastrow-Slater parameters 
(lower panels) for the optimal variational singlet wavefunctions with 
pairing wavefunction $f$ with d-wave symmetry in a 26 site cluster t-J model}\label{check26}
\end{figure}

\begin{figure}
\centerline{\psfig{figure=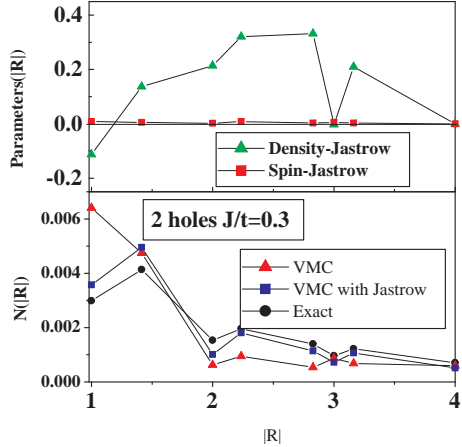,width=7cm}}
\caption{ Same as in Fig.\protect\ref{check26} 
for the 32 site cluster, with a broken symmetry 
Slater determinant wavefunction with $\Delta_{AF}=0.2t$
 $\Delta_{BCS}=0.1t$ and $t^\prime/t=-0.15$.}\label{check32}
\end{figure}

In Fig.(\ref{check26}) and Fig.(\ref{check32}) we show the full Jastrow-Slater optimization for the t-J model  
in the largest size cases  where the exact solution is known 4 holes in 26 sites\cite{poilblanc}. 
and  2 holes in 32 sites\cite{leung}, respectively,
We  display the hole-hole  correlation functions $N(R)= < (1-n_0) (1-n_R)> $ on the variational 
wavefunction with and without the Jastrow factor. We see that the improvement towards 
the exact solution is crucially dependent on the Jastrow density-density factor especially 
at long range distance. This behavior seem to be analogous to the one dimensional case where 
long range Jastrow-factors are enough to determine the anomalous long range behavior of correlation 
functions in one dimensional Luttinger liquids\cite{valenti,franjic}.
The remarkable   accuracy of the Jastrow-Slater wavefunctions is clearly limited (see e.g. Fig.\ref{nqj1}) 
to the region $J/t \lesssim 0.5$  where 
 pairs  with d-wave  symmetry repel each other and do not form many-particle bound-states 
as is the case for large $J/t$ when phase separation occurs.  

For the 32 site  cluster the spin-translation symmetry has to be explicitly broken 
in the variational wavefunction by introducing a staggered magnetization 
$H_0 \to H_0 - \Delta_{AF} \sum_R (-1)^R \sigma_R^z $ along the z axis. 
The best wavefunction compatible with reflection-rotation and translation  with spin interchange 
($\uparrow \leftrightarrow \downarrow$) can be conveniently parametrized by $\Delta_{AF}$ and a 
next neighbor hopping $t^\prime$in the Slater determinant\cite{ogata} and  also 
a spin Jastrow factor is allowed within this class of wavefunctions.
Even in this case, as shown in the corresponding Fig(\ref{check32}), the hole-hole 
correlations are almost exactly reproduced by the strong attractive Jastrow term at long 
distance. This means that in this small doping regime it is important to have a broken 
symmetry ground state, which suppress the d-wave BCS order parameter. 

The next step is to perform few Lanczos steps over these variational wavefunctions 
which  have been shown to be very  accurate but not an exact representation 
of the ground state many body wavefunction. 
In the singlet case with no broken symmetry the energy as a function of the variance 
of the energy per site:
\begin{equation} \label{variance}
{<\psi_G|(H/L)^2|\psi_G>\over <\psi_G|\psi_G>} -\left({<\psi_G|H/L|\psi_G>\over <\psi_G|\psi_G>}\right)^2
\end{equation}
is indeed smoothly related to the exact ground state energy. 
The reason is that for a good variational state the variance approaches zero as the energy 
becomes exact, by increasing the number of Lanczos iterations.\cite{bled}
\begin{figure}
\centerline{\psfig{figure=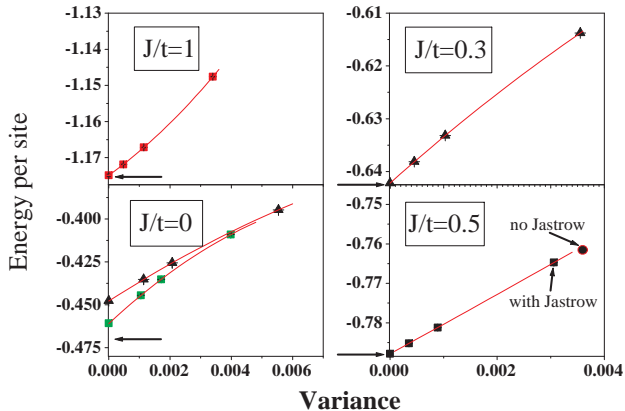,width=9cm}}
\caption{  Energy per site as a function of the number of Lanczos steps 
$p=0$ (higher energy-variance) $p=1$ (medium energy-variance) and $p=2$ (lowest energy with 
non zero variance) 
starting from the optimal Jastrow-Slater wavefunction  for  4 holes 
in the 26 site clusters and several $J/t$. Only for $J/t=0$ the broken 
symmetry solution  (with the spin Jastrow factor $v^z \ne 0$) is better than 
the best singlet wavefunction (triangles). 
The arrows indicate the exact energies, whereas 
the zero variance energies  are the extrapolated results  with a quadratic fit (continuous lines).  
For $J/t=0.5$ we show the corresponding variational $p=0$ energy and variance for the 
wavefunction without Jastrow term.}\label{ener26}
\end{figure}

\begin{figure}
\centerline{\psfig{figure=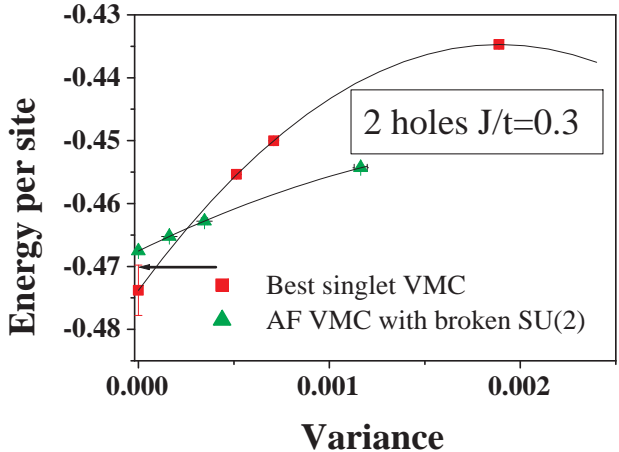,width=9cm}}
\caption{  Energy per site as a function of the number of Lanczos steps 
$p=0$ (higher energy with non zero variance) $p=1$ (medium energy with non zero variance) 
and $p=2$ (lowest energy with non zero variance) 
starting from the optimal Jastrow-Slater wavefunction for a 32-site t-J square lattice.
The arrows indicate the exact energies, whereas 
the zero variance energies  are the extrapolated results  
with a quadratic fit (continuous lines). 
 The broken 
symmetry solution is described in Fig.(\ref{check32}) 
whereas the best singlet wavefunction is obtained 
by optimizing only the density-density Jastrow  and the d-wave parameters as in 
Eq.(\ref{hbcs}).
}\label{ener32}
\end{figure}

Whenever on a finite system a broken symmetry variational 
solution has much lower energy than the fully symmetric one,   the Lanczos method 
is less effective to extrapolate  to the exact value. 
This is shown for instance for the 26 sites at $J/t=0$ Fig.(\ref{ener26}) or in the 
32 site 2 hole case Fig.(\ref{ener32}). In the latter case we show also the  energy as 
a function of the variance for the fully symmetric solution. We see in this case, that 
the approach to the exact solution for a singlet wavefunction is rather difficult but indeed possible.

On a finite system there is always a small energy gain  
 to recover  a state with definite spin.
It is very difficult to obtain this  residual energy with 
few Lanczos step iterations, since  in order 
to average over the various directions of the order parameter 
many hamiltonian power iterations are required. However this residual energy should vanish in the thermodynamic limit 
if the symmetry is indeed spontaneously broken. 
Therefore we conclude that the small discrepancy between 
exact results and extrapolated ones in Figs.(\ref{ener26},\ref{ener32}) is irrelevant within 
the above assumption, which is confirmed by simulations on much larger size, showing antiferromagnetic 
long range order at small doping\cite{calandra} and ferromagnetism for the $J=0$ case\cite{becca}.
 In any case the variance extrapolation with few Lanczos steps 
 provide always a much more accurate estimate of the exact ground state energy compared
 to the lowest  variational estimate.

\section{Results and discussion}
The variational approach is certainly 
limited and ''biased'' by the ''human'' choice 
of the  variational   wavefunction, ''believed'' to be the correct one for the 
physical problem considered. 
In this paper we have described a variational approach that improves systematically any 
given variational wavefunction with a couple (and in principle more) Lanczos steps, with a reasonable 
computational effort. This approach is certainly limited, especially for large size, when few Lanczos 
iterations cannot remove the possible large bias of the  initial variational guess.
However for 2D fermionic system 
on a lattice, in the strong correlation regime, i.e.  close to a Mott insulator state, 
it is very difficult to improve the best variational wavefunction obtained with 
the Lanczos scheme (see table). 
This result is particularly meaningful if we consider that in principle the FN technique is size consistent 
(lowers the energy per site of the variational guess even in the thermodynamic limit) and the LS technique 
with fixed number of iterations $p$ is not.
Thus, close to a Mott insulator, it is very important to change the nodes of the wavefunction -that the 
Lanczos technique allows- rather than improving only the amplitudes- as in the FN technique.
It is worth mentioning, however, that the FN energy reported in the table is only an upper bound of the expectation value 
of the hamiltonian on the FN state. We do not know exactly how much this will lower the energy 
in favor of the FN 
technique, but we expect that this should be only a minor correction, especially for large sizes.

As shown in the table the variational energy can be further improved 
by using the $r>0$ technique, 
which, within the formalism of the present paper, can be thought as a self-consistent improvement of 
the amplitudes and 
the nodes of the few Lanczos step variational wavefunction. For $r=1$ the original SR scheme 
is recovered, which as seen in Fig.(\ref{forwardfig}),  maybe not the optimal 
choice from the variational point of view.\cite{sorella,caprio}. 
As far as the variational energy is concerned the self-consistent approach ($r>0$)  is not very effective
(the improvement is between 20\% and 30\% on small size systems) 
 for small system sizes  but appears to be   more and more important as the size is increased.  
 Also correlation functions maybe qualitatively improved (Fig.\ref{nqj1})  by the self-consistent 
approach, especially when a many-particle bound state appears (''stripes'' or phase separation in this model)
which is not contained at the variational level.


An important advantage of  the standard variational approach ($r=0$) is that 
 the error in the ground state energy  and correlation functions
can be estimated  using that  the variance of the energy per site (\ref{variance})
  in an exact
calculation should be zero. The variance can be estimated  systematically
with high statistical accuracy  for  the first $p$ Lanczos steps acting
on  the initial  variational wavefunction $\psi_G$.
We show that the approach to the exact result maybe smooth, even for
large system size and number of electrons, even when $\psi_G$ is
not particularly close to the exact result.
Obvious exceptions exist
 and are   shown  here in Fig.\ref{nqj1} for correlation functions, whereas the energy 
seems always better behaved (see Fig.\ref{ener26}).    

We have tested this simple scheme  in the 2D-Heisenberg model where 
an exact solution of energies and correlation functions is easily available 
by using  standard techniques. The 2D-Heisenberg 
model has off-diagonal long-range order 
in the ground state, the order parameter 
$$m^2=  {1\over L^2} \sum\limits_{R,R^\prime} 
\vec S_R \cdot \vec S_{R^\prime} (-1)^{R-R^\prime} $$
being  finite  in the thermodynamic limit $L\to \infty$.
We start with the  variational wavefunction  in Eq.(\ref{hbcs}) 
 obtained by projecting out the doubly occupied states to a 
wavefunction with d-wave n.n. BCS pairing correlations\cite{gros}, but without 
any explicit antiferromagnetic order parameter.

\begin{figure}
\centerline{\psfig{figure=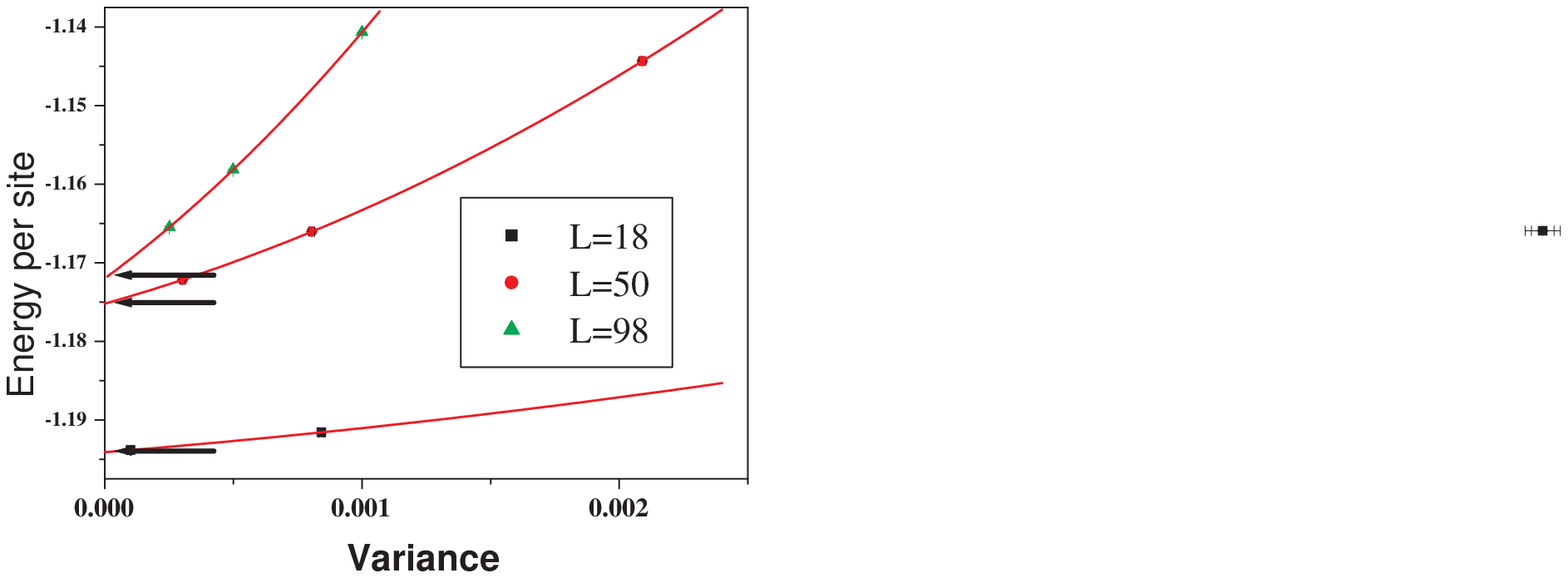,width=7cm}}
\caption{Energy per site of the finite-size Heisenberg model. 
Comparison of exact results (indicated by arrows) and the approximate $p=0,1,2$ Lanczos step
iterations  over the projected d-wave wavefunction.
Continuous lines are quadratic fit of the data.}\label{cavalloe}
\end{figure}

 This wavefunction  represents  an accurate wavefunction for the quantum antiferromagnet 
as far as  the 
energy is concerned, but certainly has not the right behavior at large 
distance,  and  maybe considered  an RVB disordered variational
 wavefunction\cite{anderson} 
After applying only two Lanczos iterations the 18 site size is almost 
exactly reproduced by this simple wavefunction, showing that at short distance 
the quantum antiferromagnetic wavefunction is almost indistinguishable 
from an RVB one.\cite{doucot} As we increase the size,
 the variational energy calculation (see Fig.\ref{cavalloe})
looses clearly accuracy, since the gap to the first excited state scales to zero 
and the Lanczos algorithm becomes correspondingly 
less effective. This loss of accuracy is however not dramatic (asymptotically 
it scales as  $~ {1\over \sqrt{L}}$), as shown by 
the similar quantitative agreement with the 
exact result  obtained both with the 50 and the 98 site lattice (Fig.\ref{cavalloe}).

\begin{figure}
\centerline{\psfig{figure=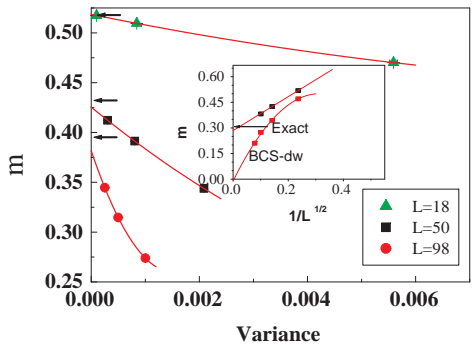,width=7cm}}
\caption{Order parameter
$m=\protect\sqrt{S(\pi,\pi)/L}$ in the finite-size 
Heisenberg model ($S(\pi,\pi)$ being the spin isotropic
antiferromagnetic structure factor). Comparison of exact results
(indicated by arrows) and the approximate $p=0,1,2$ Lanczos step
iterations over the projected d-wave wavefunction.
Continuous lines are quadratic fit of the data.  Inset: finite-size scaling
with the variational (BCS d-wave) wavefunction and with 
the variance extrapolated one.}\label{cavallom}
\end{figure}

Also remarkable  is the correct trend of long range correlation functions 
as we increase the number of Lanczos iterations. As shown in the inset of 
Fig.(\ref{cavallom})  the linear extrapolation with 
the variance, which is 
valid for  correlation functions averaged over the system volume ( see App.\ref{bulk}), 
is capable to detect almost the right long range magnetic order  in the Heisenberg model.
This is remarkable  if we consider that the starting wavefunction $\psi_G$ 
is disordered, as also shown in the same inset. 
We remark also that in this particular case the original SR algorithm\cite{sorella,caprio} 
for $r=1$ and $\Lambda \to \infty$ is {\em exact} since the nodes of the variational wavefunction
are the correct ones.

\begin{figure}
\centerline{\psfig{figure=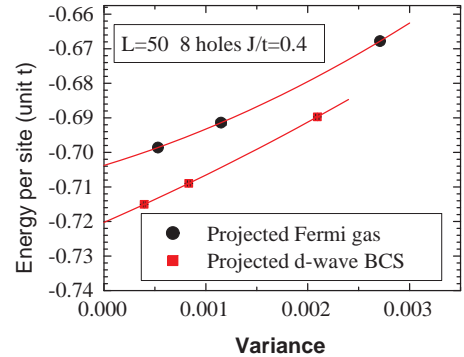,width=7cm}}
\caption{Energy per site vs. variance for two different initial wavefunctions (highest 
energy dots)  after   applying one (medium energy points) and two Lanczos steps (lowest energies points)
 Lines are quadratic fit to the data.}\label{sfiga}
\end{figure}

As anticipated the estimate of the variational  error,  by using that  
 the variance  has to approach zero in an exact calculation, 
 is really effective in this 
case, and shows that , even for large size it is possible 
to reach very good quantitative estimates of energies and correlation functions (see \cite{bled}), 
even when  the  starting wavefunction 
is not particularly close to  the exact one.
In the Lanczos algorithm the variance becomes zero only when the lowest 
energy state non orthogonal to the initial wavefunction is reached. 
However we  expect that the variance may reach very small 
values close to a ''quasi eigenstate'', implying the failure of 
the variance variational error estimate. In fact, in this case  the  energy  optimization 
Lanczos technique  is trapped in a local minimum, with 
no possibility of tunneling to the true global minimum energy.
(see fig.\ref{sfiga}).
Clearly this is a well known problem in numerical optimization  and 
the only possibility, when the exact solution is not known, is to check the 
various candidates for the ground state, and determine the one 
with the lowest energy. 
From another point of view, this property of the Lanczos algorithm maybe useful to 
estimate physically observable quantities, such as the condensation energy for 
a metal to became superconductor, which is just the macroscopic difference in energy 
between two thermodynamically stable states.
From Fig.(\ref{sfiga}) an estimate of this condensation energy  is $0.02t\sim  100K$, in reasonable 
agreement with  experiments on HTc cuprates\cite{nonloso}, 
suggesting that the main features of d-wave superconductivity can be understood with this simple model.

At the moment, the approach we have presented here   
has to be limited to very few Lanczos iterations for large size with the given 
computer resources.
Nevertheless it  is 
certainly systematic and unbiased as far as the approach to the 
ground state is concerned: the corrections to the initial guiding function 
depend only on the hamiltonian $H$ and no other biased approximation.
Compared to the standard FN technique, it allows a systematic improvement of the 
starting variational wavefunction $\psi_G$, by correcting not only its amplitudes 
but also the nodes. 

The extension of this technique to continuous models is straightforward. For the reference 
dynamic (given by $G^f$ in Eq.\ref{manusakis})  one can use the Langevin dynamic, 
as it is done  in \cite{moroni},  so that  it is possible 
to determine  the lowest energy  
state  obtained by applying to $\psi_G$ few Lanczos steps with ($r>0$) or without self-consistency($r=0$)
 or by using a many parameter variational wavefunction, with a non linear optimization 
as described in Sec.(\ref{full}).

\acknowledgements
I am deeply in debt with V. Leuveen who convinced me  in CECAM-Toulose 
about the importance to have a rigorous variational method for the ground state 
of strongly correlated systems. I am also particularly grateful to D. Poilblanc and P. W. Leung 
who kindly sent me unpublished data on the 26 and 32 site t-J model, respectively.
Special thanks to F. Becca, L. Capriotti, M. Capone , A. Parola and G. Santoro for many useful 
comments about this paper.
This work was partially supported by INFM and MURST (COFIN99). 
\appendix
\section{Properties of the operator $P_{SR}\,  ( \Lambda  I -H )\,   P_{SR}$}
\label{linear}
In this appendix we focus on some properties of the matrix $P_{SR}\,  ( \Lambda  I -H )\,   P_{SR}$:
defined in Eq.(\ref{srham}).
\begin{itemize}
\item the  maximum eigenvalue $\Lambda -E_{SR}$ of the Hermitian  matrix 
$ P_{SR} \, (\Lambda  I -H ) \, P_{SR}$
is certainly smaller than the corresponding one $\Lambda -E_0$ of the exact Green function
$\Lambda I -H$. In fact be $\psi_{SR}=P_{SR}\,  \psi_{SR}$ the  normalized 
eigenstate  of  $ P_{SR} \, (\Lambda  I -H ) \, P_{SR}$ with eigenvalue 
$E_{SR}$, then $P_{SR}^\dag=P_{SR}$ implies that 
$E_{SR}=<\psi_{SR}| P_{SR}\,  H\,   P_{SR} |\psi_{SR}>=
 <\psi_{SR}|  H  |\psi_{SR}> \ge E_0$, as 
$\psi_{SR}$ can be considered  a variational  state of the exact hamiltonian $H$ with energy 
$E_{SR}$.
\item 
 Since $\psi_{SR}$ is obtained by applying \\ 
 $\left[ P_{SR} \, (\Lambda I  -H )\, P_{SR}\right]^n$ to a given 
trial wavefunction. 
For $n\to \infty$ such  propagated wavefunction 
will converge therefore to the lowest energy state in the subspace projected by $P_{SR}$.
This implies clearly that $E_{SR}$ is 
lower or at most equal than the variational energy on the reference 
wavefunction:
$<\psi_G| H |\psi_G>$ simply because $\psi_G$ belongs to this subspace. 
\item
Since $\psi_G$ belongs to the subspace projected by $P_{SR}$, 
$<\psi_G| H \, P_{SR}= <\psi_G| P_{SR} \, H$. Therefore the  
mixed average estimate-statistically much more convenient-  
\begin{eqnarray*} 
E_{MA}&=& {  < \psi_G |  H | \psi_{SR}>  \over   
<\psi_G|\psi_{SR}> }
 = {  < \psi_G | P_{SR} \,  H \, P_{SR}| \psi_{SR}>  \over   
<\psi_G| P_{SR} | \psi_{SR}> }\\
 &=& E_{SR}
\end{eqnarray*}
, coincides with the variational bound $E_{SR}$ of the ground state energy. 
 as $\psi_{SR}$ is an exact eigenstate of $P_{SR} \,  H\,  P_{SR}$ with eigenvalue $E_{SR}$. 
\end{itemize}

\section{Forward walking} \label{forwardapp}
In order to compute correlation functions over $\psi_{SR}$ it is necessary to use a slight
generalization of the forward walking technique, generalized to non-symmetric matrix
such as (\ref{defgrefb}).
Moreover since in the meaningful SR limit of large number of walkers or bin length $L_b \to \infty$
 the parameters $\alpha_k$ can be 
assumed constants in $r_x$ (\ref{defrx}), it is much more convenient to implement the forward 
walking technique without allowing any fluctuations of the random variables $\alpha_k$.
This can be done easily by first evaluating the expectation value  ratios 
$\bar \alpha_k=<\alpha_k >/ <\alpha_0>$ $k=1,\dots p$ with the standard SR algorithm 
, i.e. allowing the $\alpha_k$ fluctuations   for each 
Markov iteration $n$. The second step is to perform  a different   simulation, usually much more 
efficient as far as the error bars are concerned, 
with $r_x$ determined by non random constants  $\bar \alpha_k$: 
\begin{equation} \label{fixrx}
r_x= 1 +\sum\limits_{k=1}^p \bar \alpha_k O^k_x
\end{equation}
If the $\bar \alpha_k$ are determined accurately for $M,L_b$ large 
 the SR conditions (\ref{srcon}) 
will be automatically verified  within error bars. 
The statistical or systematic  error related to the determination of the constants
 $\bar \alpha_k$ is also not much  important.
In fact  even assuming that 
with the first simulation the constants  $\bar \alpha_k$
are determined with a non-negligible statistical error and an unavoidable 
 systematic bias due to the 
finite number of walkers, the method that we will describe in the following will provide also in 
such a case  a variational estimate of the energy with the chosen constants 
$\bar \alpha_k$.
The analogy of this method with the Lanczos method is evident also in this case.
Even in the latter technique, a first run is usually implemented to determine 
the coefficients of the ground state $\bar \alpha_k$ in the Krilov basis spanned by the 
initial wavefunction $\psi_G $ and the powers of the hamiltonian applied to it
 $\psi_0= \psi_G+\sum\limits_{k=1,p} \alpha_k H^k |\psi_G>$ (with some 
more technical ingredient to work with an orthonormal basis). Then correlation 
functions  over $\psi_0$ 
are computed by recovering the ground state wavefunction in this basis using the 
determined coefficients $ \bar \alpha_k$.

Let us now focus on the  implementation  of the forward walking technique within the SR scheme 
at fixed constants $\bar \alpha_k$.
Since $\bar G^f$ in Eq.(\ref{defgrefb}) is not symmetric its  left eigenvector 
$<\psi_L|\propto \lim\limits_{n\to\infty} <\psi_G| ( \bar G^f)^n$ does not necessarily 
 coincide with the corresponding right eigenvector $\psi_R$.
Fortunately the matrix  $\bar G^f$ can be easily written in terms of a symmetric matrix  $G^0$
\begin{equation} \label{defgfg0}
 \bar G^f_{x^\prime,x} =\alpha_{x^\prime} G^0_{x^\prime,x} /\alpha_{x} 
\end{equation}
with
\begin{eqnarray} 
\alpha_x&=&{| \psi_G(x) ||r_x|^{r/2}  \over \sqrt{z_x} } \label{defalphax} \\
G^0_{x^\prime,x}&=&{ |r_x|^{r/2} |r_{x^\prime}|^{r/2}  \over \sqrt{z_x z_{x^\prime}} }
|\Lambda \delta_{x^\prime,x} - H_{x^\prime,x} |  \label{defg0} 
\end{eqnarray}
Therefore the right and the left eigenvectors of $\bar G^f$ are easily 
written in terms of the 
maximum eigenstate $\phi_0$ of the symmetric matrix $G_0$, namely 
$\psi_R(x)=\alpha_x \phi_0(x)$ and 
$\psi_L(x)=\phi_0(x)/\alpha(x)$. Then using the definition of the SR state 
(\ref{defpsisr}) it follows that also the left eigenvector of $\bar G^f$ can be  
 written in terms of $\psi_{SR}$:
\begin{eqnarray}
\psi_{SR}(x)&=& L_x \psi_L(x)  \label{defpsilr}  \\
{\rm with } ~~ L_x&=& \psi_G(x)  r_x/z_x   \label{deflx} \\
{\rm and } ~~ R_x&=& |r_x|^{1-r}/\psi_G(x) 
~ {\rm Sgn}\, r_x  
\end{eqnarray}

 After applying several times the Green function $\bar G^f$ 
the walkers $w^f,x$ determine the state $\psi_R(x)$. Then  
 it is possible to evaluate expectation values of 
any operator $O$ with given matrix elements 
$O_{x^\prime,x}$ by applying the following relation, which correspond to propagate $n$ times
 forward $\psi_R(x)$:
\begin{equation} \label{avsr}
{<\psi_{SR}|O|\psi_{SR}> \over <\psi_{SR}|I | \psi_{SR} > } = \lim\limits_{n\to \infty} 
{ \sum_{x^{\prime,\prime},x^{\prime},x} (\bar G^f)^n_{x^{\prime \prime},x^\prime} 
  \bar O_{x^\prime,x}  \psi_R(x) \over 
 \sum_{x^{\prime,\prime},x^{\prime},x} (\bar G^f)^n_{x^{\prime \prime},x^\prime}
 \bar I_{x^\prime,x}  \psi_R(x) }  
\end{equation}
where the matrix elements of $ O$ and the identity $I$  are replaced 
by the ones of the left-right transformed matrices $\bar O$ and $\bar I$ 
respectively. The explicit matrix elements 
of $\bar O$ and $\bar I$  in  the  RHS of the above equation are  given by:
\begin{eqnarray} 
\bar O_{x^\prime,x}  &=& L_{x^\prime}  O_{x^\prime,x} R_x \label{barme}  \\
\bar I_{x^\prime,x} &=& L_x R_x \delta_{x^\prime,x}  \label{deflxrx} 
\end{eqnarray}
This means that 
in the standard forward walking technique\cite{calandra}, instead of using 
the importance sampled matrix elements obtained with $L_x=\psi_G(x)=1/R_x$ in Eq.(\ref{barme}), 
the slightly more involved ones (\ref{barme},\ref{deflxrx}) have to be considered.
In fact by simple substitutions of these matrix elements
 into Eq.(\ref{avsr}),
 using also that $\sum_x^{\prime,\prime} (\bar G^f)^n_{x^{\prime,\prime},x^\prime} \propto 
\psi_L(x^\prime)=\psi_{SR}(x^\prime) / L_{x^\prime} $ (\ref{defpsilr}) 
and that $\psi_R(x)= \psi_{SR}(x)/R_x $  (\ref{defpsisr}), 
 Eq.(\ref{avsr}) is easily verified.
The statistical algorithm used to evaluate the ratio in
 Eq.(\ref{avsr}) is very similar 
to the standard ''forward walking'' technique\cite{calandra} 
for diagonal operators.
The few differences are:
\begin{itemize} 
\item also the denominator in Eq.(\ref{avsr}) has to be ''forward''
 propagated for $n$ iterations, since in this 
case the diagonal elements of $\bar I$  are not trivially one (since $L_x\ne R_x^{-1}$).
The error bars have to be then calculated taking into account  that the numerator and the 
denominator are very much correlated.
\item off diagonal operators can be computed without performing another 
simulation, 
provided the matrix elements of the operator $O$ are contained in the non vanishing ones 
of the Green function $G$ (or some power of $G$ if the operator is evaluated 
statistically).
In particular the expectation value of the hamiltonian and the even more
 accurate ones (\ref{defpowers}) can be computed  altogether 
 with  a single Markov chain.
\item similarly the accuracy of diagonal and off diagonal operators can be
 further improved by 
computing  
$$ <\psi_{SR}| G^k O G^k |\psi_{SR}> \over <\psi_{SR}| G^{2k} |\psi_{SR}>.$$
In fact an important advantage of the SR technique is that the reference 
Green function 
$G^f_{x^\prime,x}$ is non zero for all non-zero elements of the exact Green
 function $G$, (whereas in the FN technique the matrix elements  with
 negative sign are suppressed).
 Thus the exact sampling of the Green function $G$ can be done with the
 standard reweighting method,  requiring only  the 
finite multiplicative factors 
$s_{x^\prime,x} =G_{x^\prime,x}/G^f_{x^\prime,x}$, 
calculated for each iteration $n$ and each walker 
of  the Markov chain. 
The same technique can be obviously generalized when the reference Green function $G^f$ is 
the fixed node one-slightly generalized to have the possibility to cross the nodes\cite{caprio}-, 
simply replacing $G^f$ and $z_x=1$ in the above expressions.
However the statistical accuracy for the determination of the constants $\left\{\alpha_k\right\}$ is 
vary bad with the FN reference $G^f$, about an order of magnitude less efficient than the Eq.(\ref{manusakis}) one,
 without a significant improvement in variational energies. 
The reason of such bad   behavior (or the successful one for 
\ref{manusakis}) is not clear at present.

\end{itemize}

\section{Efficient calculation of the single Lanczos step wavefunction}
\label{onelan}
In this Appendix we describe an efficient way to find the optimal 
LS wavefunction $|\psi_{\alpha}\rangle=(1+\alpha H)|\psi\rangle$, 
starting from a chosen variational guess $|\psi\rangle$, {\em i.e.},
to calculate the value of $\alpha$ for which the energy
\begin{equation}\label{ealpha}
E(\alpha)=\frac{\langle\psi_G|(1+\alpha H)H(1+\alpha H)|\psi_G\rangle}
{\langle\psi_G|(1+\alpha H)^2|\psi_G\rangle}
\end{equation}
has a minimum. 
A standard method is to calculate statistically the various powers of the hamiltonian 
\begin{equation} \label{momenta}
h_n= { \langle \psi_G | H^n | \psi_G \rangle \over \langle \psi_G  | \psi_G \rangle } ,
\end{equation}
using  configurations $x$  generated by the Metropolis 
algorithm according to the weight $\psi_G(x)^2$.
This method is however inefficient since much better  importance sampling 
is obtained when configurations are instead generated according to the optimal 
Lanczos wavefunction $\psi_{\alpha^*}(x)=\left(1+\alpha^* e_{\psi_G} (x) \right) \psi_G(x)$,
where 
$e_{\psi}(x)={\langle \psi | H |x \rangle \over \langle \psi |x \rangle }$\cite{notezero}
is the local energy corresponding to a generic guiding wavefunction $\psi$,  and $\alpha^*$  minimizes the above expectation value (\ref{ealpha}) 
for $\alpha=\alpha^*$.
This wavefunction $\psi_{\alpha^*}$ maybe 
much better leading to much lower variances 
especially for the higher momenta $h_2$ and $h_3$.

In this Appendix we describe an efficient way to find the optimal 
LS wavefunction $|\psi_{\alpha^*}\rangle$, 
starting from a chosen variational guess $|\psi_\alpha \rangle$ with energy:
\begin{equation} \label{ealphan}
E(\alpha) = { h_1 +2 \alpha h_2 + \alpha^2 h_3  \over 1 + 2 \alpha h_1 + \alpha^2 h_2 }
\end{equation}
easily written in terms of the energy momenta $h_n$.
 
In order to minimize (\ref{ealphan}), given an arbitrary value of $\alpha$, it is convenient  
first to compute the energy expectation value $h_1$ with the standard statistical method 
and then, in place of the remaining hamiltonian higher momenta $h_2$ and $h_3$, 
 generate statistically configurations according to $\psi_{\alpha} (x)^2$ and compute:
\begin{eqnarray}\
E(\alpha) &=& \frac{\langle\psi_{\alpha}|H|\psi_{\alpha}\rangle}{\langle\psi_{\alpha}|\psi_{\alpha}\rangle}  \nonumber \\
\chi&=&\frac{\langle\psi_{\alpha}|(1+\alpha H)^{-1}|\psi_{\alpha}\rangle}{\langle\psi_{\alpha}|\psi_{\alpha}\rangle}~,\nonumber 
\end{eqnarray}
$E(\alpha)$ is obtained by averaging over the chosen configurations the local energy  
corresponding to $\psi_{\alpha}$, namely    $< e_{\psi_{\alpha}} >$ whereas $\chi$ is obtained by averaging 
over the same  configurations $ < (1+ \alpha e_{\psi_G} (x) )^{-1} >$.
Given $\chi$  it is straightforward to compute $$h_2=[(\chi^{-1} -2) (1+\alpha h_1)+1]/\alpha^2$$, 
and therefore given $h_1$ and $h_2$, the value of $E(\alpha)$ implicitly defines the highest 
momentum $h_3={ E(\alpha) (1+2 \alpha h_1 +\alpha^2 h_2 ) -h_1 -2 \alpha h_2 \over \alpha^2 }$.
Notice that  the most difficult energy momentum $h_3$ is given by sampling an energy expectation value, 
which is by far statistically more  accurate compared to the direct  determination of $h_3$.

It is then possible to minimize analytically $E(\alpha)$, yielding:  
\begin{equation} \label{alpha*}
\alpha^*={ -(h_3-h_1 h_2) \pm \sqrt{ (h_3 -h_1 h_2)^2 -4 (h_2-h_1^2) (h_1 h_3 -h_2^2) } \over 2 ( h_1 h_3 -h_2^2)  }
\end{equation}
where the above sign $\pm$  is such to  minimize $E(\alpha^*)$.

 The analytic minimization of $E(\alpha)$ (\ref{ealphan}), given the values of $\chi$,   $h_1$ and $E(\alpha)$ 
itself, provides the exact  value of $\alpha^*$ in   Eq. (\ref{alpha*})  within the statistical uncertainties.
They  become smaller and smaller whenever $\alpha \sim \alpha^*$.
Typically two or at most three attempts are enough to reach an accurate determination of $\alpha^*$
when the condition:
\begin{equation} \label{minlan}
\chi= {1 \over 1 + \alpha^* E(\alpha^*) }
\end{equation}
is exactly fulfilled.
This condition is true in general only for the eigenstates of the Hamiltonian, 
but remains valid  for the single Lanczos step wavefunction.

\section{Variance estimate of the error on ''bulk`` correlation functions} \label{bulk}
In this appendix we  estimate the error on correlation functions 
assuming that the ground state $|\psi_0>$ is approximated with the 
wavefunction $\psi_p>$  distant $\epsilon_p$ from $|\psi_0>$  
Namely. with no loss of generality we write:
\begin{equation}\label{defeps}
|\psi_0>=|\psi_p>+\epsilon_p |\psi^\prime>
\end{equation}
with $<\psi_p|\psi_p>=<\psi^\prime|\psi^\prime>=1$, $\psi^\prime$, 
representing a normalized wavefunction orthogonal to the exact one 
$<\psi_0|\psi^\prime>=0$.
We restrict our analysis to thermodynamically averaged correlation 
functions $O$, the ones which can be written as a bulk average of local 
operators $O_R$: $O={1\over L} \sum_R O_R$.
This class of operators include for instance the average
 kinetic or potential energy or the spin-spin correlation function 
at a given distance $\tau$> $O_R= S_R \cdot S_{R+\tau} $.
 If we use periodic 
boundary conditions the expectation value of $O_R$ 
on a state with given momentum  does not even depend on $R$ and the  
bulk average  does not represents an approximation 
\begin{equation}\label{defc} 
{<\psi_0 | O_R |\psi_0> \over <\psi_0 |\psi_0> } = 
{<\psi_0 | O |\psi_0> \over <\psi_0 |\psi_0> } = C .
\end{equation}

 We  show here that  the  expectation value of bulk 
averaged operators $O$ on the approximate 
state $\psi_p$ satisfy the following relation:  
\begin{equation}\label{approx}   
<\psi_p |O|\psi_p>= C+ O(\epsilon_p^2,\epsilon_p/\sqrt{L})
\end{equation}
thus implying that for large enough size the  
expectation value (\ref{approx})  approaches  the 
exact correlation function  $C$ linearly with the variance.
 This allows to obtain a good accuracy with a good variational calculation, 
that  is not  easy to obtain if a term  $\sim \epsilon_p$ dominates.
 
 The validity of  the above statement is  very simple to show under 
very general grounds.
In fact by definition:
\begin{equation}
<\psi_p|O|\psi_p> = C+ 2 \epsilon_p <\psi^\prime|O|\psi_0> + \epsilon_p^2 
<\psi^\prime|O |\psi^\prime> 
\end{equation}
The term proportional to $\epsilon$ in the above equation can be 
easily bounded by use of the Schwartz inequality:
\begin{equation}
|<\psi^\prime|O|\psi_0>|^2 = |<\psi^\prime|O-C |\psi_0>|^2 \le 
<\psi_0|(O-C)^2 |\psi_0> 
\end{equation}
The final term in the latter inequality can be estimated  
under the general assumption that correlation functions
 $C(\tau)= <(O_R-C)  (O_{R+\tau}-C) >$ decay sufficiently fast 
with distance $|\tau|$, as a consequence of the cluster property:
$$<\psi_0|(O-C)^2 |\psi_0> = (1+\epsilon^2)  {1\over L} \sum_\tau C(\tau).$$
This concludes the proof of the statement of this appendix, provided 
$\sum_\tau C(\tau)$ is finite for $L\to \infty$.

\clearpage

\begin{table}
\label{table}
\begin{tabular}{ccccccccccc}
$N$ & $L$  & J/t & VMC & VMC+LS & VMC+FN  & VMC+2 LS & VMC+LS +FN & Best SR & Best r &  Exact \\
\hline\hline
22  & 26  &  0.3   &  -0.6138(1)    &   -0.6332(1) & -0.6277(1)    &  -0.6381(1)  & -0.6371(1)   & -0.6387(1) & 0.375  & -0.64262  \\
22  & 26  &  0.5   &  -0.7647(1)    &   -0.7812(1) & -0.7759(1)    &  -0.7852(1)  & -0.7841(1)   & -0.7855(1) & 0.25&  -0.78812  \\
22  & 26  &  1.0   &  -1.1476(1)    &   -1.1672(1) & -1.1608(1)    &  -1.1719(1)  & -1.1706(1)   & -1.1724(1)& 0.25 & -1.17493  \\
30  & 32  &  0.3   &  -0.4543(1)    &   -0.4628(1) & -0.4611(1)    &  -0.46522(3) & -0.46524(3)  & -0.4661(1)& 0.375  &   -0.470175  \\
84  & 98  &  0.4   &  -0.6653(1)    &   -0.6807(1) & -0.6777(1)    &  -0.6865(1)  & -0.68530(5) & -0.6879(2) & 0.1 & -0.692(1)  \\
50  & 98  &  0.4   &  -0.9656(1)    &   -0.9832(1) & -0.98225(5)   &  -0.9886(1)  & -0.98781(6) & -0.9901(2) &  0.1  & -0.9920(5)  \\
\hline\hline
\end{tabular}
\caption{Energy per site in the t-J model for various  variational methods.
VMC is the standard variational method, VMC+LS is obtained by applying to it a 
Lanczos step, VMC+FN is the lattice fixed node approach\protect\cite{bemmel,ceperley1}, 
VMC+2 LS  indicate  the 
two Lanczos step variational wavefunction, VMC+LS+FN, is the fixed node 
over the VMC+LS wavefunction, and the ''Best r'' is $<psi_{SR}|H|\psi_{SR}>$, computed by ''forward walking'' 
as described in App.(\ref{forwardapp}), by optimizing the parameter $r$.  
The exact energy values for the largest size was estimated by the variance extrapolation.
On the 98 sites,the FNLS computation 
takes 10 hour CPU time on a Pentium-II 400MHz, whereas the VMC+2LS wavefunction takes about 40 hours 
with $M=500$ walkers for a statistical accuracy of $10^-4t$ on the energy per site, the ''best SR'' 
another factor  $8$ more due to the forward walking.
 The computation of diagonal correlation functions  instead takes a similar amount of time for all the methods, 
thus it is safer to compute them with the best variational method.Error bars are indicated in brackets.}
\end{table}

\begin{references}
\bibitem{onebody} In this paper we adopt a more general definition for 
 a Slater determinant, as any state which is an eigenstate of a one body 
hamiltonian, not necessarily commuting with the particle number.
\bibitem{ineq} This is strictly valid provided $G$ is a positive definite operator, 
which generally holds for large enough $\Lambda$.
In fact by the Schwartz inequality: it is simple to show that 
 the expectation value of the energy on 
$P_{SR} \psi_{n+1}$ is lower or at most equal than the one over $P_{SR} \psi_n$.
\bibitem{notezero} If there exist 
configurations $x$ such that $\psi_G(x)=0$ and $\psi_G^\prime(x)\ne 0$, it is 
computationally convenient to regularize the given guiding function  by 
adding to it a random noise which maybe arbitrary small  within computer accuracy, thus 
preventing the above pathology with probability one. By reducing this noise to zero 
the ''clean'' limit  can be easily achieved without too much effort or complicated 
changes in the algorithm.\cite{manusakis}
\end{references}
\end{document}